\documentclass[twocolumn,twocolappendix]{aastex}
\usepackage{amsmath,amssymb}
\usepackage{times}
\usepackage[utf8]{inputenc}
\usepackage{booktabs}
\usepackage[nolist,nohyperlinks]{acronym}
\usepackage{xspace}
\usepackage{breakurl}
\usepackage{multirow}
\usepackage{threeparttable}
\usepackage{ulem}

\def\nsbh{\ac{NS}-\ac{BH}\xspace}
\def\bns{\ac{NS}-\ac{NS}\xspace}
\def\joint{\ac{GW}-\ac{SGRB}\xspace}
\newcommand{\mBH}{M_{\rm BH}}
\newcommand{\mNS}{M_{\rm NS}}
\newcommand{\mbNS}{M_{\rm b, NS}}
\newcommand{\rNS}{R_{\rm NS}}
\newcommand{\spin}{\chi_{\rm BH}}
\newcommand{\mRem}{M_{\rm rem}}
\newcommand{\mDisk}{M_{\rm disk}}
\newcommand{\EISO}{E_{\gamma, {\rm ISO}}}
\newcommand{\EISOobs}{E_{\gamma, {\rm ISO}}^{\rm obs}}
\newcommand{\thj}{\theta_{\rm j}}

\definecolor{darkpastelgreen}{rgb}{0.01, 0.75, 0.24}

\shorttitle{NS Radius from GW and SGRB NS-BH Observations}

\begin{document}

\title{Constraining the Neutron Star Radius with Joint
  Gravitational-Wave and Short Gamma-Ray Burst Observations of Neutron
  Star--Black Hole Coalescing Binaries}

\date{\today}

\author{Stefano Ascenzi}%
\affil{%
  Universit\`a di Roma Tor Vergata, Via della Ricerca Scientifica 1,
  I-00133 Roma, Italy \&
  Sezione INFN di Roma Tor Vergata, I-00133 Roma, Italy\\
  INAF - Osservatorio Astronomico di Roma, Monte Porzio Catone (RM),
  Italy\\%
  Dip. di Fisica, Universit\`a di Roma  ``Sapienza,'' P.le A. Moro, 2,
  I-00185 Rome, Italy}
\author{Nicola De Lillo}%
\affil{%
  Universit\`a di Trento, Dipartimento di Fisica, Via Sommarive 14,
  I-38123 Povo, Trento, Italy%
}
\author{Carl-Johan Haster}%
\affil{%
  Canadian Institute for Theoretical Astrophysics, University of
  Toronto, Toronto, Ontario M5S 3H8, Canada%
}
\author{Frank Ohme}%
\affil{%
  Max-Planck-Institut f\"{u}r Gravitationsphysik,
  Albert-Einstein-Institut, Callinstraße 38, 30167 Hannover, Germany\\
  Leibniz Universit\"at Hannover, 30167 Hannover, Germany
}
\author{Francesco Pannarale}%
\affil{%
  Gravity Exploration Institute, School of Physics and Astronomy, Cardiff University, The Parade, Cardiff CF24 3AA, UK\\
  Dipartimento di Fisica, Universit\`a di Roma ``Sapienza,'' Piazzale A.\,Moro 5, I-00185, Rome, Italy\\
  INFN Sezione Roma1, Piazzale A.\,Moro 5, I-00185, Rome, Italy\\
  \href{mailto:francesco.pannarale@ligo.org}{francesco.pannarale@ligo.org}
}

\begin{abstract}
  Coalescing \nsbh binaries are promising sources of \acp{GW} that are predicted to be
  detected within the next few years by current \ac{GW}
  observatories. If the \ac{NS} is tidally disrupted outside the
  \ac{BH} innermost stable circular orbit, an accretion torus may
  form, and this could eventually power a \ac{SGRB}. The observation
  of an \ac{SGRB} in coincidence with gravitational radiation from an
  \nsbh coalescence would confirm the association between the two
  phenomena and also give us new insights into \ac{NS} physics. We
  present here a new method to measure \ac{NS} radii and thus
  constrain the \ac{NS} \acl{EOS} using joint \ac{SGRB} and \ac{GW}
  observations of \nsbh mergers. We show that in the event of a joint
  detection with a realistic \ac{GW} \ac{SNR} of $10$, the \ac{NS}
  radius can be constrained to $\lesssim 20$\% accuracy at 90\%
  confidence.
\end{abstract}



\maketitle

\begin{acronym}
\acrodef{BH}[BH]{black hole}
\acrodef{EM}[EM]{electromagnetic}
\acrodef{EOS}[EOS]{equation of state}
\acrodef{GRB}[GRB]{gamma-ray burst}
\acrodef{GW}[GW]{gravitational-wave}
\acrodef{ISCO}[ISCO]{innermost stable circular orbit}
\acrodef{KAGRA}[KAGRA]{Kamioka Gravitational wave detector}
\acrodef{NS}[NS]{neutron star}
\acrodef{SGRB}[SGRB]{short gamma-ray burst}
\acrodef{ShortGRB}[SGRB]{Short gamma-ray burst}
\acrodef{SNR}[S/N]{signal-to-noise ratio}
\end{acronym}

\section{Introduction}
\acresetall


The first observation of a binary \ac{BH} merger in \acp{GW} made by
Advanced LIGO, GW150914, marked the dawn of the \ac{GW} astronomy
era~\citep{GW150914}.  Subsequently, the LIGO-Virgo Collaboration
reported other nine binary \ac{BH} merger
observations~\citep{GW151226, GW170104, GW170608, GW170814,
  LIGOScientific:2018mvr},
and the detection of GW170817, a signal that is consistent with a
binary \ac{NS} inspiral~\citep{GW170817}.  
\citet{Hinderer2018} showed that \nsbh systems with certain
parameter combinations are also consistent with the \ac{GW} and \ac{EM}
observations of GW170817.

Second-generation \ac{GW} detectors --- {\it i.e.}, Advanced
LIGO~\citep{AdvLIGO}, Virgo~\citep{AdvVirgo}, KAGRA~\citep{KAGRA}, and
LIGO-India~\citep{M1100296, LigoIndia} --- will also be able to detect
the \ac{GW} radiation emitted by \nsbh coalescing binaries, a category
of compact binary that remains to be observed.  In addition to
\acp{GW}, among the reasons of interest in coalescing \nsbh binaries
is the possibility that if the \ac{NS} is tidally disrupted outside
the \ac{ISCO} of its \ac{BH} companion, matter can be accreted onto
the \ac{BH}, powering a \ac{SGRB}~\citep{Nakar:2007yr}.  We now know
that a binary \ac{NS} merger can power an
\ac{SGRB}~\citep{GW170817GRB}, and future joint \ac{GW}-\ac{EM}
observations will be able to determine whether this is true for \nsbh
systems too.  Naturally, such observations are intrinsically
challenging due to the low expected \joint joint detection rate for
\nsbh binaries.  This is predicted by \citet{C15} to be
$0.4$--$10\,\mathrm{yr^{-1}}$ for LIGO-Virgo at design sensitivity and
an idealized \ac{SGRB} observing facility with all-sky coverage, in
line with earlier results from \citet{Nissanke:2012dj} (up to
$3\,\mathrm{yr^{-1}}$ with a three detector network when ignoring
source inclination requirements).  The estimate drops to
$0.03$--$0.7\,\mathrm{yr^{-1}}$ when considering the \emph{Swift}
field of view.  For comparison, \citet{Wanderman:2014eza} calculated
joint detection rates with {\it Swift} and {\it Fermi} of
$0.3$--$1.4\,\mathrm{yr^{-1}}$ and $3$--$10\,\mathrm{yr^{-1}}$,
respectively, while \citet{Regimbau:2014nxa} determined
$0.001$--$0.16\,\mathrm{yr^{-1}}$ in the case of {\it Swift}.  The
assumptions behind these frameworks are different and we refer the
interested reader to the original articles for details.  The upcoming
third generation of \ac{GW} detectors, however, will have a much
larger observational horizon (up to $z\simeq 4$ for \nsbh binaries)
which automatically increases the joint detection rate considerably
\citep{Punturo:2010zz, ETdesignstudy,  Kalogera:2019sui,
  Sathyaprakash:2019rom}.  Further interest in \nsbh binaries is due
to the possibility that the tidally disrupted material is ejected away
from the \nsbh system, generating an \ac{EM} transient powered by the
decay of $r$-process ions (macronova)~\citep{LP98, Kulkarni:2005jw,
  Metzger10, MB12, Fernandez:2015use, Metzger2017}.  Similar to the
\ac{SGRB} case, recent \ac{GW}-\ac{EM} observations of GW170817 have
confirmed that binary \acp{NS} are sites that host
$r$-processes~\citep{GW170817MMA, GW170817kn}, but whether this holds
for \nsbh binaries as well, remains to be proven observationally.

Whether the \ac{NS} in an \nsbh binary undergoes tidal disruption or
not, and the amount of matter that is available for accretion (or to
feed into the ejecta) in the event of a tidal disruption, both depend
on the physical properties of the \ac{BH} (mass and spin) and of the
\ac{NS}, including the currently unknown \ac{EOS} that regulates the
microphysics of the  \ac{NS}~\citep{Pannarale2010, Foucart2012,
  Foucart2018}.  The \ac{GW} radiation of coalescing \nsbh systems
also depends on the source properties, and among them is the \ac{NS}
\ac{EOS}~\citep{BC92, KS95, V00, Shibata:2009cn, DFKOT10, Kyutoku2010,
  KOST11, Lackey2014, LK14, Foucart:2012vn, Foucart:2014nda,
  Pannarale2013a, Pannarale2015a, Pannarale2015b, Kawaguchi:2015bwa,
  HTF16, KPP17, DK18}, but it may be hard to constrain the
  \ac{NS} \ac{EOS} with \nsbh \ac{GW} inspiral signals
  only~\citep{PannaraleRezzolla2011}.  Therefore, the \ac{GW} and
\ac{EM} emission of \nsbh binaries that undergo tidal disruption will
carry information about all the properties of the progenitor system,
and hence about the \ac{NS} \ac{EOS}.

\citet{Pannarale:2014rea} showed how joint \ac{GW} and \ac{SGRB}
observations of \nsbh coalescences may provide invaluable information
about the \ac{NS} \ac{EOS}.  On the basis of this observation, we
propose a method to exploit such observations in order to constrain
the \ac{NS} radius, and thus the \ac{NS} \ac{EOS}.  In the scenario in
which \nsbh systems are progenitors of \ac{SGRB} central engines, it
is reasonable to expect the \ac{SGRB} energy to be proportional to the
rest mass of the torus that accretes onto the remnant \ac{BH}.  In
turn, this mass can be expressed as a function of the mass and spin of
the \ac{BH} initially present in the binary, and the \ac{NS} mass and
radius~\citep{Foucart2012, Foucart2018}.  Our method explores the
portion of parameter space that is pinpointed by the \ac{GW}
observation --- \ac{GW} Bayesian inference provides posterior
distributions for the two masses and the \ac{BH} spin --- and thus
determines a posterior distribution for the \ac{NS} radius by imposing
the condition that the merger yields a torus sufficiently massive to
power the observed \ac{SGRB} energy.

Assuming an \ac{SGRB} isotropic energy of $\EISO = 10^{51}\,$erg, we
expect to be able to measure the \ac{NS} radius (at 90\% confidence)
with $\lesssim 20$\% accuracy, given a \ac{GW} detection with a \ac{SNR}
of $10$. This measure is expected to improve for less energetic
\acp{SGRB} and \acp{GW} with higher \ac{SNR}.  We show that the poorly
known parameters that our analysis marginalizes over --- such as the
mass-energy conversion efficiency for the \ac{SGRB} --- have a
negligible impact on our results, provided the \ac{SGRB} energy is
sufficiently low.  Our method is well behaved even for (non-isotropic)
energies as high as $E_\gamma = 10^{50}\,$erg, thus the restriction is
not very limiting.

The paper is organized as follows. In Sec.\,\ref{sec:methodology} we
describe our method in detail, discussing the poorly constrained
parameters involved in the analysis.  In Sec.\,\ref{sec:results} we
test the method and present the results we obtained by simulating
joint \joint observations. Finally, in
Sec.\,\ref{sec:discussion} we draw our conclusions.

Throughout the paper, we assume geometric units ($G=c=1$), unless
otherwise explicitly noted.

\section{Methodology}
\label{sec:methodology}
When an \ac{NS} undergoes tidal disruption during an \nsbh
coalescence, part of the matter that constitutes it may remain outside
the \ac{BH} up to a few milliseconds after the merger.  We denote the
mass of this remnant matter by $\mRem$.  A small fraction of this will
form unbound ejecta that can eventually power \ac{EM} transients by
radioactive decay of $r$-process heavy ions~\citep{LP98,
  Kulkarni:2005jw, Metzger10, MB12, Fernandez:2015use, Metzger2017}.
The rest of it will stay bound around the \ac{BH}, forming a neutrino-cooled accretion disk and a tidal tail, orbiting with high
eccentricity, which will fall back, filling the disk on a timescale of
$0.1$--$1$\,s~\citep{Foucart2012}.
The remnant \ac{BH} and the disk form a system that is a plausible
candidate for the central engine of (a fraction of) \acp{SGRB}, as the
accretion of mass from the disk onto the \ac{BH} could power the
launch of a relativistic jet~\citep{ELPS89, P91,MR92, NPP92,  Mes06,
  LRR07}.

Given a disk of mass $\mDisk$, the energy radiated in gamma rays
during the prompt emission by conversion of mass corresponds to
\begin{equation}
  E_\gamma=\epsilon \mDisk \,, 
  \label{eq:mDiskEgamma}
\end{equation}
where $\epsilon$ is the mass-energy conversion efficiency.  $E_\gamma$
is related to the \ac{SGRB} isotropic energy $\EISO$ by
\begin{equation}
  E_\gamma=(1-\cos{\thj})\EISO
\end{equation}
where $\thj$ is the jet half-opening angle, {\it i.e.} its beaming
angle\footnote{This expression holds for a simple, top-hat jet model.
  It can be replaced with a more complicated angle dependency that
  appropriately models a structured jet.}.  In this work, we assume
$\EISO$ to be measured from the gamma-ray flux, provided the distance
to the host galaxy of the \ac{SGRB} is known.  We may therefore write
\begin{equation}
  (1-\cos{\thj})\EISO=\epsilon \mDisk \,. 
  \label{eq:EnergyMass}
\end{equation}
Assuming the gravitational radiation emitted by the coalescence is
also observed, one can exploit this last equation to connect the
measured $\EISO$ and the \nsbh properties inferred from the \ac{GW}
measurement (masses and spins of the binary constituents, as discussed
later on in this section) in order to constrain the \ac{NS} radius,
and hence the \ac{NS} \ac{EOS}.

Two unknowns are evident in Eq.\,(\ref{eq:EnergyMass}).  The first one
is the efficiency $\epsilon$, which varies from system to system and
is determined by a chain of complicated physical processes, the nature
of which is an open field of investigation (see, {\it e.g.},
\citet{Nakar:2007yr} and \citet{LRR07}, and references therein).  The treatment
of $\epsilon$ in our analysis is discussed in
Sec.\,\ref{sec:epsilonPrior}.  The second unknown is the beaming angle
$\thj$.  While this can be inferred by measuring the time at which a
jet break appears in the afterglow light curve~\citep{SPH99}, usually
\ac{SGRB} jet breaks are not observed and only lower limits ($\thj
\gtrsim 3^\circ$) can be placed~\citep{Berger2014}.  This happens
because (i) \ac{SGRB} afterglows are fainter than long
\acused{GRB}\ac{GRB} afterglows, and because (ii) their light curves
typically drop below a detectable level within a day.  We therefore
treat $\thj$ as an unknown parameter in our analysis, as detailed
further in Sec.\,\ref{sec:thetaPrior}.

The last element entering Eq.\,(\ref{eq:EnergyMass}) is the disk mass
$\mDisk$, and we make the approximation $\mDisk\simeq \mRem$ ({\it
  i.e.}, we neglect the mass of the possible ejecta\footnote{The
  observation of the kilonova emission from the same event, or the
  lack thereof, could be used to constrain the ejecta mass, and
  therefore to assess the systematics deriving from this
  approximation.}).  This approximation is justified by the results of
numerical-relativity simulations, which predict ejecta masses of at
most $\sim \mathcal{O}(10^{-2} \, M_\odot)$~\citep{Kawaguchi:2015bwa,
  KI15, KKST16, Foucart2017} and total remnant masses that are an
order of magnitude higher in such extreme cases~\citep{KOST11,
  Foucart2012, Foucart2017}.
  
We express $\mRem$ using the semi-analytical formula of
\citet{Foucart2018}, which updates a formula previously introduced
in~\citet{Foucart2012} and is obtained by fitting results of
fully relativistic numerical-relativity simulations.  Specifically,
the fraction of \ac{NS} matter that remains outside the remnant
\ac{BH} is given by
\begin{equation}
  \frac{\mRem}{\mbNS}=\Bigl[\alpha \frac{1-2C_{\rm NS}}{\eta^{1/3}}-\beta\hat{R}_{\rm ISCO}\frac{C_{NS}}{\eta} + \gamma\Bigr]^\delta\,,
  \label{eq:Foucart}
\end{equation}
where $\mbNS$ is the baryonic mass of the \ac{NS}, $\eta =
\mBH\mNS/(\mBH+\mNS)^2$ is the symmetric mass ratio ($\mBH$ and $\mNS$
being the gravitational mass of the \ac{BH} and the \ac{NS},
respectively), $\rNS$ is the radius of the \ac{NS} at isolation
expressed in Schwarzschild coordinates, $C_{\rm NS}=\mNS/\rNS$ is the
\ac{NS} compactness, $\spin$ is the dimensionless spin magnitude of
the \ac{BH} in the \nsbh binary, $\hat{R}_{\rm ISCO}=R_{\rm
  ISCO}/\mBH$ is the normalized \ac{ISCO} radius, and $\alpha=0.406$,
$\beta=0.139$, $\gamma =0.255$, $\delta =1.761$ are the free
coefficients determined by the fitting procedure.\footnote{We omit the
  $\max$ between $0$ and the term in square brackets of
  Eq.\,(\ref{eq:Foucart}) that appears in the original expression for
  $\frac{\mRem}{\mbNS}$ given in \citet{Foucart2018}.  The reason for
  this is explained in Sec.\,\ref{sec:results}.}  The \ac{ISCO} radius
$R_{\rm ISCO}$ is a function of the mass $\mBH$ and spin magnitude
$\spin$ of the \ac{BH} in the original \nsbh
binary~\citep{Bardeen1972}.

The discussion carried out so far can be summarized as follows: an
\nsbh coalescence can result in an \ac{SGRB} with energy proportional
to the rest mass liberated by the tidal disruption and given by
Eq.\,(\ref{eq:Foucart}).  The system of equations laid out is closed
by prescribing an \ac{EOS} for the \ac{NS}.  This enters the
expression(s) for the remnant mass through $\rNS$ and $\mbNS$.  Given
that our goal is to determine a method to constrain the \ac{NS}
\ac{EOS} on the basis of a joint \joint observation of an \nsbh
coalescence, the \ac{EOS} is ultimately the unknown we would want to
solve for, under the constraints imposed by the observational data.
In order to simplify this task and to avoid repeatedly solving the
Tolman-Openheimer-Volkoff \ac{NS} structure equations \citep{OV39, T39}, we express the \ac{NS} baryonic mass $\mbNS$ as a function of
the \ac{NS} gravitational mass $\mNS$ and solve for $\rNS$.  In this
sense, our method constrains the \ac{NS} radius and indirectly
constrains the \ac{NS} \ac{EOS}.

\begin{figure}
  \includegraphics[width=\columnwidth]{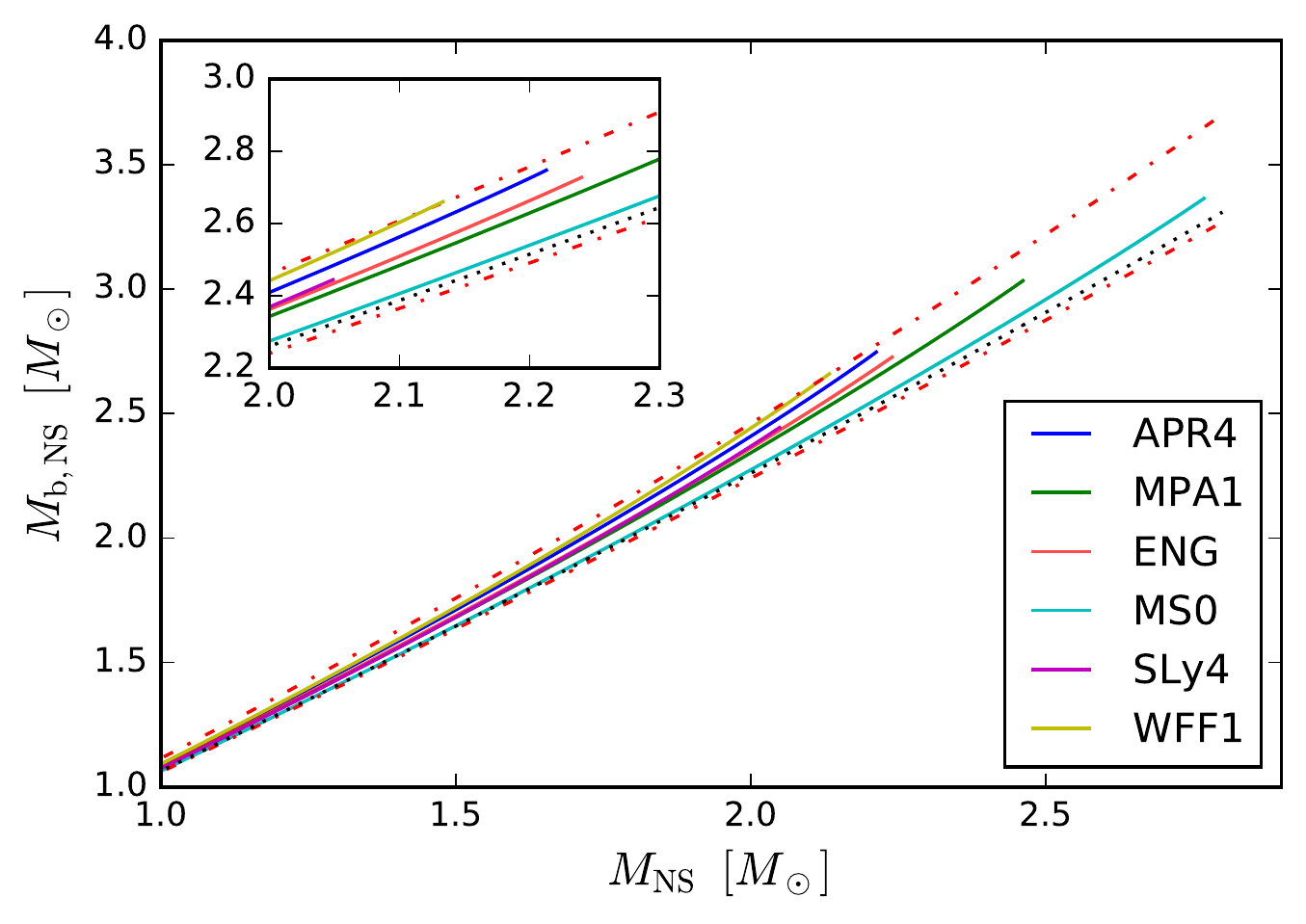}
  \caption{Baryonic-gravitational mass relations along stable \ac{NS}
    equilibrium sequences obtained with different \acp{EOS}
    (continuous curves). The black dotted line is the fit in
    Eq.\,(\ref{eq:mbNS}) with its original value
    $c_2=13/200$~\citet{Cipolletta2015}.  The red dotted-dashed lines
    correspond instead to $c_2 = \{12/200, 23/200\}$ (lower and upper
    curve, respectively).  These two values allow us to enclose all
    the \ac{NS} equilibrium sequences.}
  \label{fig:c2}
\end{figure}

The approximation we use to relate $\mNS$ to $\mbNS$ is the fit to
\ac{NS} equilibrium sequences provided by~\citet{Cipolletta2015}:
\begin{equation}
  \frac{{M}_{\rm b, NS}}{M_\odot}=\frac{{M}_{\rm NS}}{M_\odot}+c_2\left(\frac{M_{\rm NS}}{M_\odot}\right)^2\,.
  \label{eq:mbNS}
\end{equation}
The value of the free coefficient $c_2=13/200$ found by Cipolletta and
collaborators is biased by their choice of \acp{EOS} used to build the
\ac{NS} equilibrium sequences they fit with Eq.\,(\ref{eq:mbNS}).  We
find that, for a large sample of \acp{EOS}, acceptable values of $c_2$
lie in the range $[12/200, 23/200]$ as shown in Fig.\,\ref{fig:c2},
where we only show six representative \acp{EOS} to avoid overcrowding
the figure. 

Given the observation of an \nsbh coalescence, \ac{GW} parameter
estimation provides posterior probability distributions for the
gravitational masses and the \ac{BH} spin that enter
Eq.\,(\ref{eq:Foucart}).  
Once we obtain the raw posterior distribution samples from the \ac{GW}
analysis, we ``prune'' them as follows.  We discard all parameter
points that do not satisfy the requirements $M_1>3\,M_\odot$ ({\it
  i.e.}, the primary object is presumably not a \ac{BH} because it is
not massive enough), $M_2<2.8\,M_\odot$ and $\chi_2<0.4$ ({\it i.e.},
the secondary object is presumably not an \ac{NS} because its mass
and/or spin are too high) \footnote{This is the only step where the information on the spin of the secondary object is exploited.}.  This step allows us to downsample the
posteriors of the \ac{GW} measurement to a set of points reasonably
compatible with the assumption that the observed \ac{SGRB} was due to
an \nsbh progenitor.


After the pruning of the \ac{GW} posteriors,
we determine a posterior for the \ac{NS} radius $\rNS$ as follows.  We
randomly sample the joint \ac{GW} posterior distribution for $\mNS$,
$\mBH$ and $\spin$ (effectively using it as an informed prior in a
hierarchical analysis), and assume uniform prior distributions for
$\rNS$ and the remaining unknowns in our setup, {\it i.e.},
$\epsilon$, $c_2$ and $\cos\thj$.  From Eq.\,(\ref{eq:EnergyMass}) we
thus obtain a distribution for $\EISO$.  Each value of this
distribution is then compared to $\EISOobs$, which is the measured
value of $\EISO$.
We then reject any sample point that yields an energy that differs by
more than a given tolerance $\tau$ from the observed $\EISO$,
according to the condition
\begin{equation}
  \frac{|\EISO - \EISOobs|}{\EISOobs} > \tau
  \label{eq:tolerance}
\end{equation}
Here $\tau$ accounts both for an uncertainty on the observed \ac{SGRB}
energy and for errors introduced by using the approximate formula in
Eq.\,(\ref{eq:Foucart}), which \citet{Foucart2018} reported to be $\sim 15\%$.
We also reject any sample point that yields a violation of the causality condition $\rNS \ge 3.04\, G \mNS/c^2$ \citep{Lattimer1990, Glendenning1992}.

It is important to stress that although the \ac{GW} signal alone can bring information on $\rNS$ ({\it e.g.}, via the tidal deformability of the star), this information is not exploited by our method, in order to keep our analysis as agnostic as  possible. While this constitutes a loss of information, we avoid introducing systematic errors due to the modeling of the \ac{EOS} imprints on the \ac{GW} signal waveform.  We will explore the benefits of
using the full \ac{GW} information in a future study.

The building blocks of our method are summarized as follows:
\begin{itemize}
    \item Parameter estimation on the \ac{GW} signal to obtain posterior distribution on the signal parameters. Among these, $M_1$ (mass of the primary star), $M_2$ (mass of the secondary star), and $\chi_1$ (spin of the primary star) are those that will directly enter in our analysis. An example of this is provided in Section \ref{sec:recovery}.
    \item The posterior distribution of \ac{GW} signal parameters is ``pruned'' to reject all the parameter combinations incompatible with a \nsbh system. The criterion involves conditions on $M_1$ , $M_2$, and $\chi_2$ (spin of the secondary star) described earlier in this section. After this step we can set $M_1 = \mBH$,  $M_2 = \mNS$ and $\chi_1 = \spin$.
    \item We use the distributions of $\mNS$, $\mBH$ and $\spin$ as \ac{GW} informed priors. We define priors for the other parameters $\thj$, $\epsilon$, $c_2$, and $\rNS$ (an example of this is reported in Section \ref{sec:Priors}). Sampling $N$ times over the joint distribution of $\mNS$, $\mBH$, $\spin$, $\thj$, $\epsilon$, $c_2$ and $\rNS$, we can solve Eq.\,(\ref{eq:EnergyMass}) [with $\mRem$ provided by Eq.\,(\ref{eq:Foucart})] for each sample point in order to obtain a posterior distribution on $\EISO$.
    \item We reject all sample points that do not satisfy the condition in Eq.\,(\ref{eq:tolerance}) as well as the causality constraint in order to obtain a posterior distribution on $\rNS$.
    
\end{itemize}

The next section presents the application of our method to
several simulated joint \ac{GW}-\ac{SGRB} observations.



\section{Method performance assessment}
\label{sec:results}
\subsection{Injection of the signal}
To assess the performance of our method, we simulate various joint
\joint observations of \nsbh coalescing binaries characterized by the
sets of parameters reported in Table \ref{tab:cases}.  The ``true''
reference value of the \ac{NS} radius --- {\it i.e.}, the quantity
that our method aims at recovering --- 
is determined solving Eq.\,(\ref{eq:Foucart}) for $\rNS$, once the parameters $\mNS$, $\mBH$,
$\spin$, and $E_{\gamma, ISO}$ of the simulated observation are
specified:

\begin{equation}
  \rNS=\frac{\bigl(2\alpha\,\eta^{-1/3}+\beta \hat{R}_{\rm ISCO}\,\eta^{-1}\bigr)\mNS}{\alpha\,\eta^{-1/3} + \gamma-[E_\gamma/(\epsilon \mbNS)]^{1/\delta}}\,.
  \label{eq:reverseFouc}
\end{equation}
Here we also substituted $E_\gamma/\epsilon$ for $\mRem$.

The three remaining free parameters are set to $c_2 =
17/200$, $\cos\thj=0.98$, {\it i.e.}, $\thj\simeq 11^\circ$, and
$\epsilon = 0.01$, which are all within their respective prior
distribution ranges.  These choices do not affect the final outcome of
our analysis, but only serve the purpose of providing a target value
for the \ac{NS} radius.

The
properties of the simulated \nsbh coalescences are given in Table
\ref{tab:cases} with masses specified in their respective source
frame, the \ac{BH} spin $\spin$ being aligned with the orbital angular
momentum and assuming the \ac{NS} is non-spinning.  We also assume
alignment between the total angular momentum and the line of sight,
consistent with an observation of both \acp{GW} and an \ac{SGRB} jet.
To highlight the capabilities of the analysis presented in this paper,
and to remove sources of both systematic and statistical
uncertainties, the \ac{GW} signal is injected into a data stream
without added Gaussian noise, and both the injected signal and the
parameter estimation analysis are using the \textsc{IMRPhenomPv2}
\ac{GW} model~\citep{Hannam:2013oca, Khan:2015jqa, Husa:2015iqa,  
  Smith:2016qas}.  This model includes an effective treatment of the
spin-precession dynamics, but does not take the imprint of possible
\ac{NS} tidal disruptions onto the \ac{GW} signal into account.  Thus,
the $\rNS$ constraints presented in this study can be taken as lower
bounds, as further direct information about the \ac{NS} properties
should only act to narrow these constraints.

As reported in Table \ref{tab:cases}, for each of the three \nsbh
systems we consider, we use two values of the isotropic energy.  This
allows us to assess how this quantity affects the measurement of
$\rNS$.  We inject the \nsbh \ac{GW} signals at two values of
\ac{SNR}, namely, $30$ and $10$, 
which correspond to sources at redshift $z\simeq 0.04$ and $z\simeq
0.12$, respectively.

\begin{table}[tb]
  \resizebox{\columnwidth}{!}{%
    \begin{tabular}{{c@{\hspace{0.3cm}}c@{\hspace{0.3cm}}c@{\hspace{0.3cm}}c@{\hspace{0.3cm}}c@{\hspace{0.3cm}}c@{\hspace{0.3cm}}c@{\hspace{0.3cm}}}}
      \toprule[1pt]
      \toprule[1pt]
      \addlinespace[0.3em]
      \multirow{2}{*}{Label} & $\mNS$ & $\mBH$ & \multirow{2}{*}{$\spin$} & $\EISOobs$ & $\rNS$ & $\mathcal{M}_{\rm c}$ \\
      & $[M_\odot]$ & $[M_\odot]$ & & $[10^{50}\,{\rm erg}]$ & $[{\rm km}]$  & $[M_\odot]$ \\
      \addlinespace[0.2em]
      \midrule
      \addlinespace[0.2em]
      \texttt{m484chi048L} &1.35 & 4.84 & 0.48 & $1$  & 10.124 & 2.14 \\
      \texttt{m484chi048H} &1.35 & 4.84 & 0.48 & $50$ & 10.521 & 2.14\\
      \texttt{m484chi080L} &1.35 & 4.84 & 0.80 & $1$  &  7.797 & 2.14\\
      \texttt{m484chi080H} &1.35 & 4.84 & 0.80 & $50$ &  8.103 & 2.14\\
      \texttt{m100chi070L} &1.35 & 10.0 & 0.70 & $1$  & 11.183 & 2.93\\
      \texttt{m100chi070H} &1.35 & 10.0 & 0.70 & $50$ & 11.569 & 2.93\\
      \bottomrule[1pt]
      \bottomrule[1pt]
    \end{tabular}
  } 
  \caption{Parameters describing the joint \joint observation scenarios considered in this work. Each case is labeled by the \ac{BH} mass and spin, while the last letter refers to the \ac{SGRB} (simulated) observed isotropic energy (L/H for low/high).  The \ac{NS} radius $\rNS$ is determined from Eq.\,(\ref{eq:reverseFouc}) after setting $c_2 = 17/200$, $\cos\thj=0.98$, and $\epsilon = 0.01$. All masses are defined in their respective source frame.}
   \label{tab:cases}
\end{table}

\subsection{Recovery of the GW signal and parameter estimation}
\label{sec:recovery}
The parameter estimation of the \ac{GW} signal is performed using the 
\textsc{LALInference} package~\citep{Veitch2015,lalinference_o2}
assuming a detector network consisting of LIGO--Hanford and
LIGO--Livingston, both operating at their nominal design
sensitivities~\citep{Aasi:2013wya, TheLIGOScientific:2014jea}.

In the parameter estimation analysis we perform an ``agnostic'' recovery, where we assume a prior distribution on
the detector-frame masses as uniform within $[1.0, 14.3] \,M_\odot$
with additional constraints on both the (gravitational) mass ratio $[1
\leq \mNS/\mBH \leq 1/8]$ and chirp mass, $\mathcal{M}_c = (\mBH
\mNS)^{3/5}(\mBH+\mNS)^{-1/5}$, within $[2.18, 4.02] \,M_\odot$.  We
allow for isotropically distributed spins with dimensionless
spin magnitudes of $[0 \leq \chi \leq 0.89]$ for both binary objects,
but as the injected binary is viewed face-on we expect only a minimal
information contribution from the binaries'
spin-precession~\citep{Fairhurst:2017mvj}.  The analysis assumes a
uniform-in-volume distribution for the sources' luminosity distance,
and because we require a joint \joint observation we assume the
direction of the \ac{SGRB} as known and fix the sky location to its
true values in the \ac{GW} analysis.  Finally, we allow for
isotropically oriented binaries, with no restrictions on the binary
inclination or constraints from the allowed beaming angles in the
\ac{GW} analysis itself.
The results of the parameter estimation on the \ac{GW} injected signals are 
summarized in Table \ref{tab:fullposterior} in the Appendix, where the $90\%$ of credible intervals on the masses, the spin of the primary star, and the mass ratio $q$ are reported. In Table \ref{tab:nsbh_posterior} we also reported on the $90\%$ intervals for the same quantities obtained after the pruning of the posteriors.


\subsection{Prior distributions for the remaining parameters}
\label{sec:Priors}
As discussed in Section \ref{sec:methodology}, values of the parameters $\epsilon$, $\thj$,  $c_2$ and $\rNS$ must be provided in order to solve Eq.\,(\ref{eq:EnergyMass}) to obtain $\EISO$. These are sampled from the prior distributions defined in this dedicated Section.
\subsubsection{Prior distribution for $\epsilon$ \label{sec:epsilonPrior}}
\label{sec:epsilon}
The efficiency $\epsilon$ introduced in Eq.\,(\ref{eq:mDiskEgamma}) is
poorly constrained.  It can be expressed as the product of
$\epsilon_{\rm jet}$, which is the efficiency of conversion of
accreted rest mass into jet kinetic energy, and $\epsilon_{\gamma}$,
which is the conversion efficiency from jet kinetic energy to
gamma-ray radiation.  \citet{Zhang2007} measured the latter efficiency
for a sample of long and short \emph{Swift} \acp{GRB} finding values
between $30\%$ and $60\%$, with an average of $49\%$.  The efficiency
$\epsilon_{\rm jet}$ is not directly measurable and depends on the
nature of the jet launching mechanism.  This can be driven by
magnetohydrodynamics~\citep{BZ77, BP82, PGB15} or by
neutrino-antineutrino pair annihilation~\citep{ELPS89, ZB11}. In both
cases its value depends upon the spin of the remnant
\ac{BH}~\citep{ZB11, PGB15}.  In a context similar to ours,
\citet{Giacomazzo2013} used a value of $\epsilon=\epsilon_\gamma\times
\epsilon_{\rm jet}=0.05$.  In our analysis, we draw random values of
$\epsilon$ according to a uniform prior distribution between $0$ and
$0.2$ [according to \citet{LRR07} it is unlikely for mass to be
converted into energy with an efficiency higher than $\sim 0.1$].
%

It is worth noting that, at a given an energy $E_\gamma$, there is a
degeneracy between the \ac{NS} radius and $\epsilon$.  Physically, one
can think of the system being able to increase/decrease $E_\gamma$ by
increasing/decreasing its $\epsilon$ or $\mDisk$.  The latter may in
turn be obtained with an increase/decrease in $\rNS$.  To understand
how a specific $\epsilon$ may affect the inferred value of $\rNS$, we
refer to Eq.\,(\ref{eq:reverseFouc}), where we can see that $\rNS$ is roughly independent
of $\epsilon$ for
$\epsilon\gg E_\gamma/\mbNS$.  If we consider an \ac{NS} with $\mbNS\sim
1.5\,M_\odot$, powering \acp{SGRB} with energies
$E_\gamma=\{10^{49},10^{50}, 10^{51}\}\,$erg would require
efficiencies $\epsilon\gg \{10^{-4},10^{-3}, 10^{-2}\}$ in order for
the inferred value of $\rNS$ to not be significantly affected.  These
efficiency values are at most of the same order of magnitude as the
ones inferred for the magnetohydrodynamics mechanisms considered
in~\citet{HK06} and~\citet{PGB15}, which
inspired~\citet{Giacomazzo2013} to adopt the fiducial value of
$\epsilon=5\%$.  The efficiency for the neutrino-antineutrino
annihilation mechanism is expected to be lower, in general, but values
of the same order as for the magnetohydrodynamics mechanisms have been
found for high \ac{BH} spins and mass accretion rates~\citep{SRJ04,
  ZB11}.  Nevertheless, in order to power an \ac{SGRB} with a remnant
mass value up to $\mathcal{O}(0.1\,M_\odot)$ \citep{KOST11,
  Foucart2012, Foucart2017}, the efficiency cannot be lower than
$10^{-6}$.  Thus, the dependency of $\rNS$ on $\epsilon$ is expected
to be weak for faint events even in the case of neutrino-antineutrino
pair annihilation.

Finally, if $\EISO\lesssim 10^{50}\,$erg, the dependency of $\rNS$ on
the beaming angle and $c_2$ is also weak, because the term
\begin{equation}
  \frac{E_\gamma}{\epsilon \mbNS}=\frac{(1-\cos{\thj})\EISO}{\epsilon (M_{\rm NS}+c_2M^2_{\rm NS}/M_\odot)}
\end{equation}
in the denominator of Eq.\,(\ref{eq:reverseFouc}) becomes negligible.
Therefore, in this circumstance, our results will not depend on the
particular prior distribution choices for $c_2$ and $\thj$.

\subsubsection{Prior distribution for $\thj$ \label{sec:thetaPrior}}
The information about \ac{SGRB} beaming angles is sparser than that for long \acp{GRB}.  The~\citet{Berger2014} review, for example,
reported a mean beaming angle of $\langle \thj \rangle \gtrsim
10^\circ$ for \acp{SGRB} and clearly shows how this angle is measured
only in a handful of cases.  The maximum measured value of $\thj$ is
about $25^\circ$, which was obtained in a single instance.  In this
work, we therefore consider a cosine-flat prior distribution for
$\thj$, with angle values limited to the range $[1^\circ, 30^\circ]$.
However, we note that additional \ac{EM} follow-up observations of a
specific \nsbh coalescence event and its host galaxy could potentially
further constrain the sampling interval for $\thj$.
Finally, it is worth noticing that, concerning the \ac{GW} side, it is unlikely to measure the inclination of the binary system with a precision that allows us to constrain $\thj$ (assuming it is less than $\sim 50^\circ$)~\citep{Fairhurst:2017mvj}.

\subsubsection{Prior distributions for $\rNS$ and $c_2$}
While the \ac{NS} \ac{EOS} binds together the values of $\rNS$ and
$c_2$ at a fundamental level, we use a simplified setup in which both
(unknown) quantities are sampled from two independent uniform prior
distributions.  Our uniform prior distribution for the \ac{NS} radius
runs from 9 to 15\,km.  This range encompasses the known limits on
\ac{NS} radii that come from observational and theoretical constraints
[for reviews on this topic, see \citet{OF16} and \citet{LP16}], as
well as the limits inferred from the analysis of the tidal effects of
GW170817~\citep{GW170817EOS}.  As stated previously, we found that
Eq.\,(\ref{eq:mbNS}) can accommodate a large set of \ac{NS}
equilibrium sequences built upon different \acp{EOS}, provided that
$c_2$ is allowed to vary between $12/200$ and $23/200$.  In order to
be as agnostic as possible about the \ac{EOS} of \ac{NS} matter, we
adopt a uniform distribution for the unknown $c_2$ over such an interval.
The impact of this prior on our results is negligible, which lends
support the our simplification of sampling $c_2$ and $\rNS$
independently. This is due to the fact that $c_2$ enters
Eq.\,(\ref{eq:reverseFouc}) via the \ac{NS} baryonic mass ${M}_{\rm b,
  NS}$ (see\,Eq.\,\ref{eq:mbNS}) in a term that is of the form
$\epsilon c_2 \mNS^2$.  This term is clearly dominated by the prior on
$\epsilon$, which is a truly unknown parameter, and $\mNS$, which is
constrained by the \ac{GW} analysis.

\subsection{Results}


Given the results of the \ac{GW} parameter estimation analysis and the pruning of these results to account only for \nsbh systems, we
sample $N$ points\footnote{Typically, we set $N=3\times 10^{6}$ for
  cases with $\EISO = 10^{50}\,$erg and $N=10^5$ for cases with
  $\EISO=5\times 10^{51}\,$erg.} of the mass and spin \emph{pruned}
posterior distributions to obtain parameters that we feed into
Eq.\,(\ref{eq:EnergyMass}), which we then solve for $\EISO$ (under the
$\mDisk\simeq\mRem$ approximation in Sec.\,\ref{sec:methodology}).
Eq. (\ref{eq:Foucart}) can be used to determine $\mRem$ as a
function of the \nsbh parameters.

Once this step is complete, each of the $N$ sample points of the
(pruned) \ac{GW} posterior is associated with a value of $\EISO$.  We
can then use the condition given in Eq.\,(\ref{eq:tolerance}) with
$\tau\equiv 2$ to determine the subset of sample points with
combinations of parameters such that the energy $\EISO$ they return
lies within a 200\% relative difference from the observed energy
$\EISOobs$.  The absolute value that appears in
Eq.\,(\ref{eq:tolerance}) allows for combinations of the parameters
$\mBH$, $\mNS$, and $\spin$ that yield a non-physical remnant mass and
hence a non-physical $\EISO$.  Accepting non-physical remnant masses
--- rather than setting the hard cut $\mRem=0$ present in the original
formulation of \citet{Foucart2018} whenever Eq.\,(\ref{eq:Foucart})
yields a non-physical value --- corresponds to introducing an
uncertainty on the $\mRem=0$ boundary pinpointed by the fitting
formula for $\mRem$.
%

\begin{figure}[!t]
  \centering
  \includegraphics[width=\columnwidth]{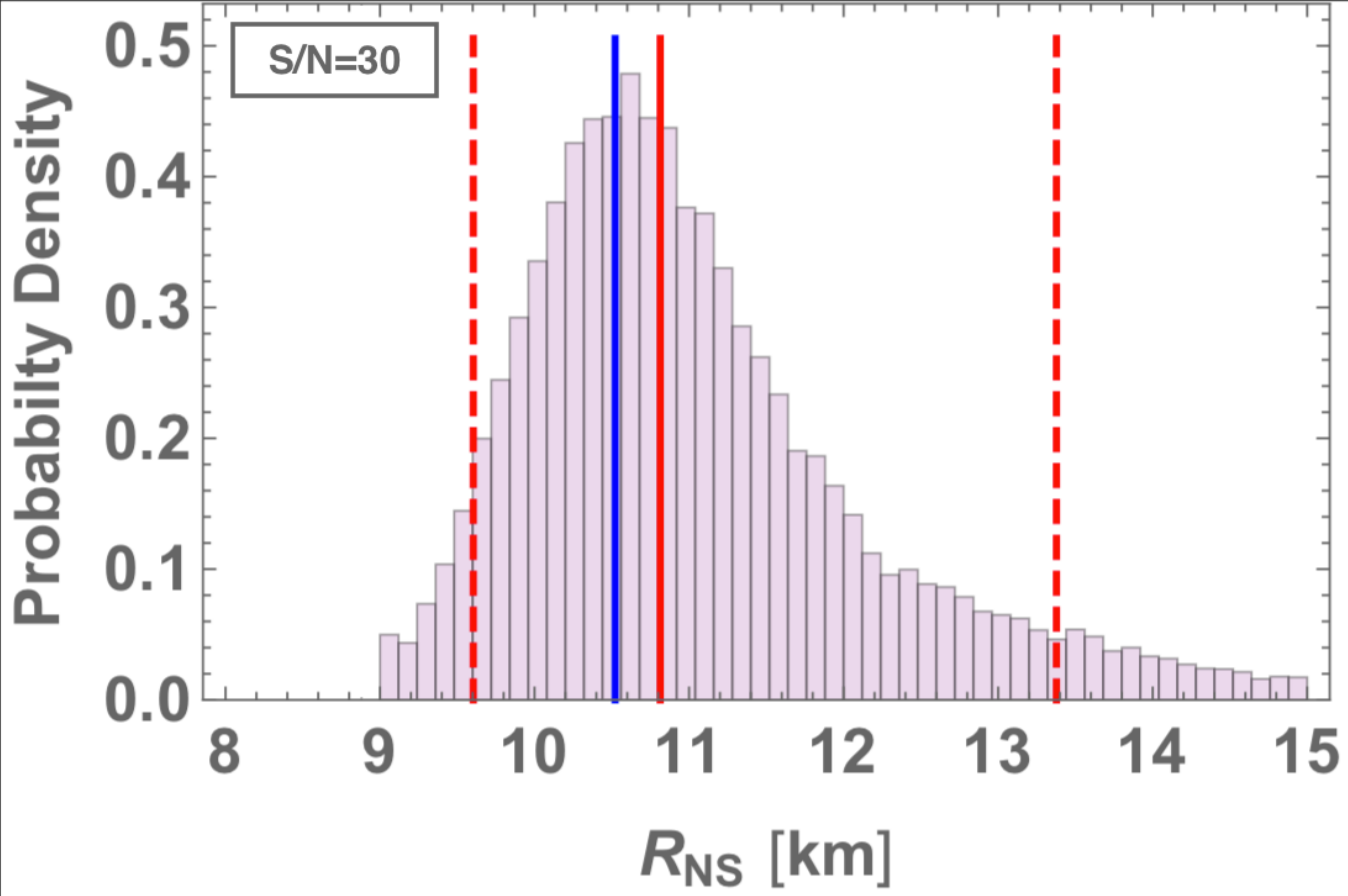}\\
  \includegraphics[width=\columnwidth]{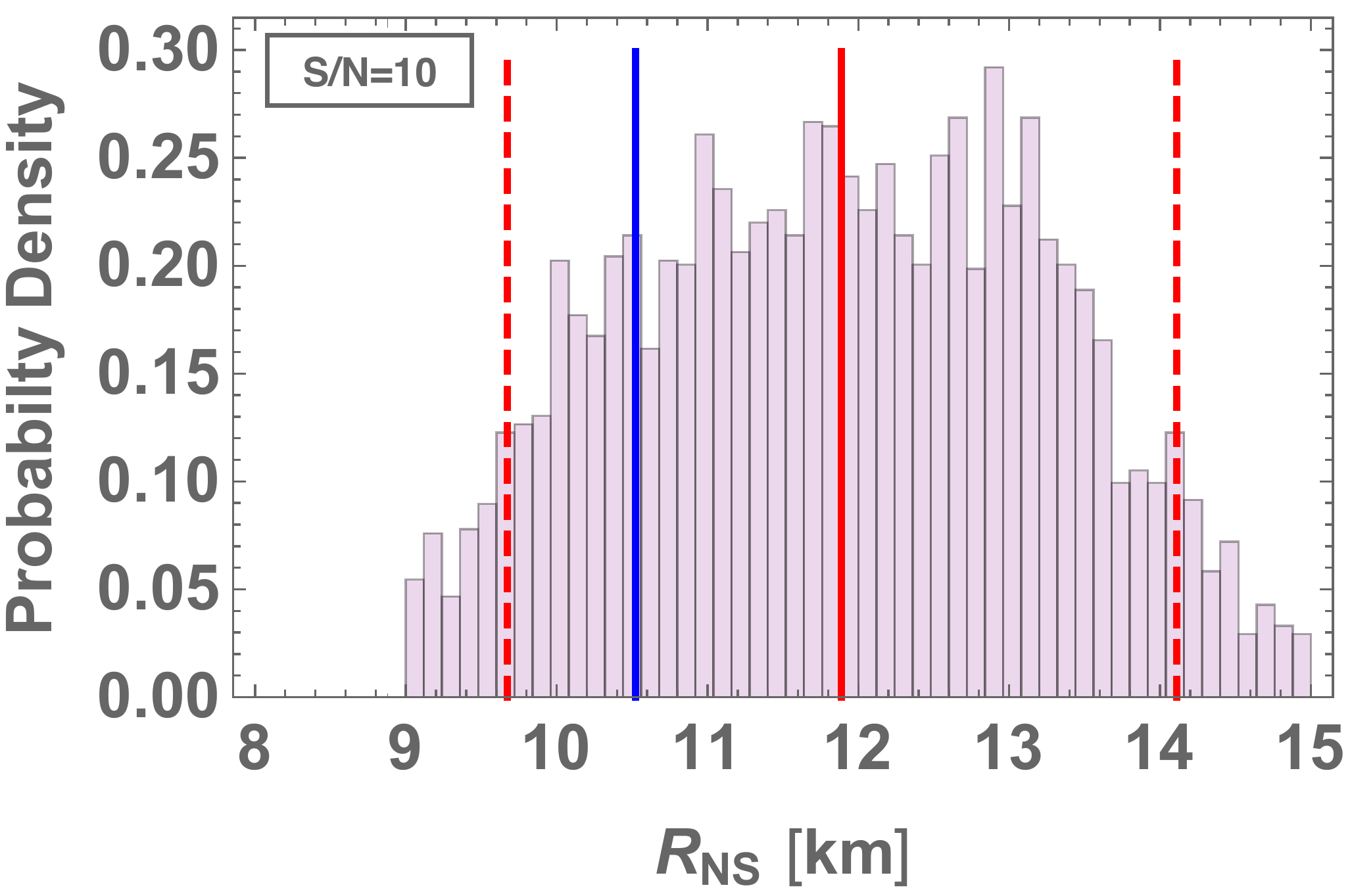}
  \caption{$\rNS$ posterior distribution for
    case \texttt{m484chi048H} at ${\rm\ac{SNR}}=30$ (top panel) and
    ${\rm\ac{SNR}}=10$ (bottom panel). The red line indicates the median value of the
    posterior, while the red dashed lines mark the 90\% credible
    interval.  The blue line represents the injected value of the
    radius.}
  \label{fig:HistoResult}
\end{figure}

\begin{figure}[!t]
  \centering
  \includegraphics[width=\columnwidth]{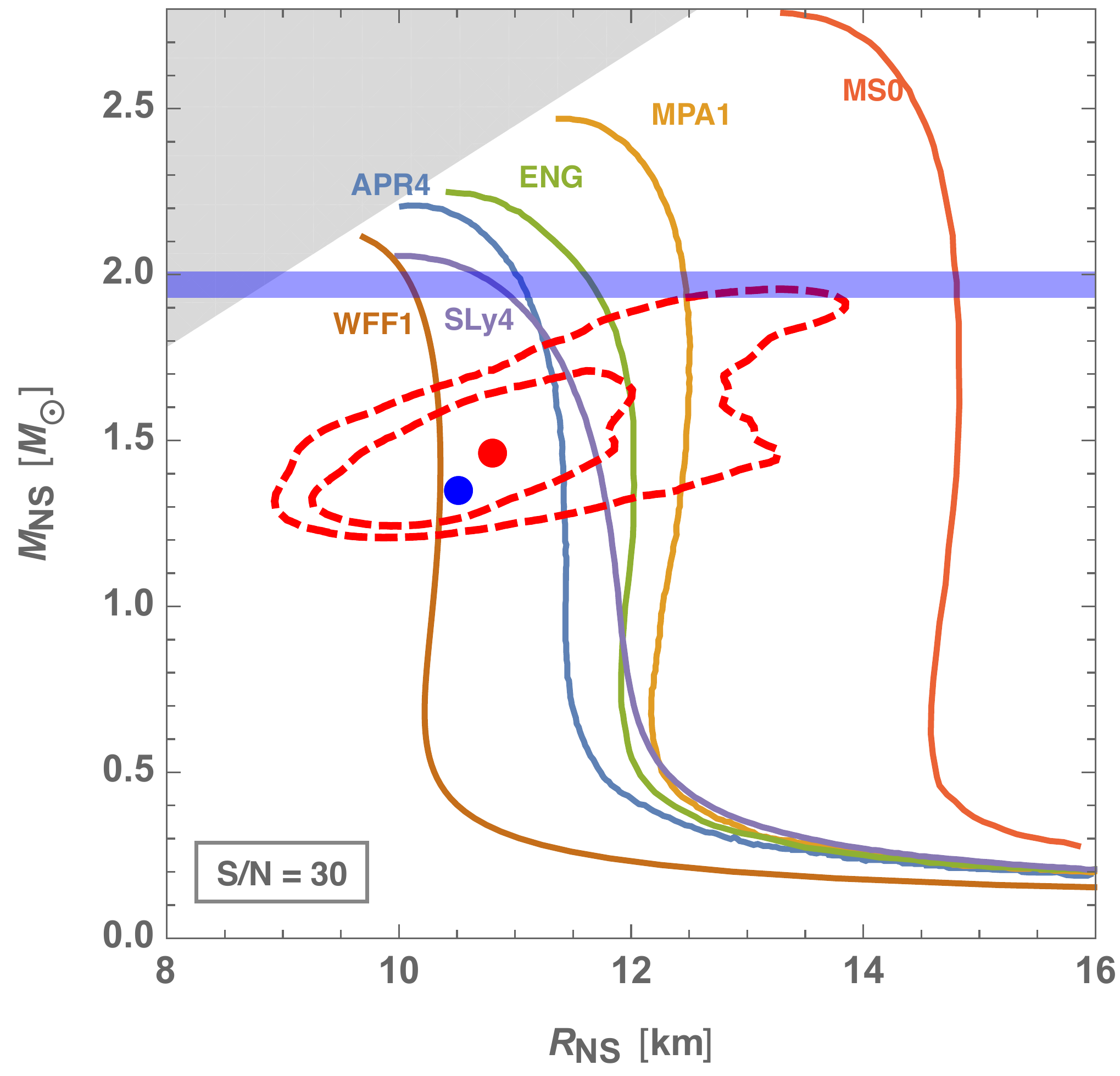}\\
  \vspace{0.2cm}
  \includegraphics[width=\columnwidth]{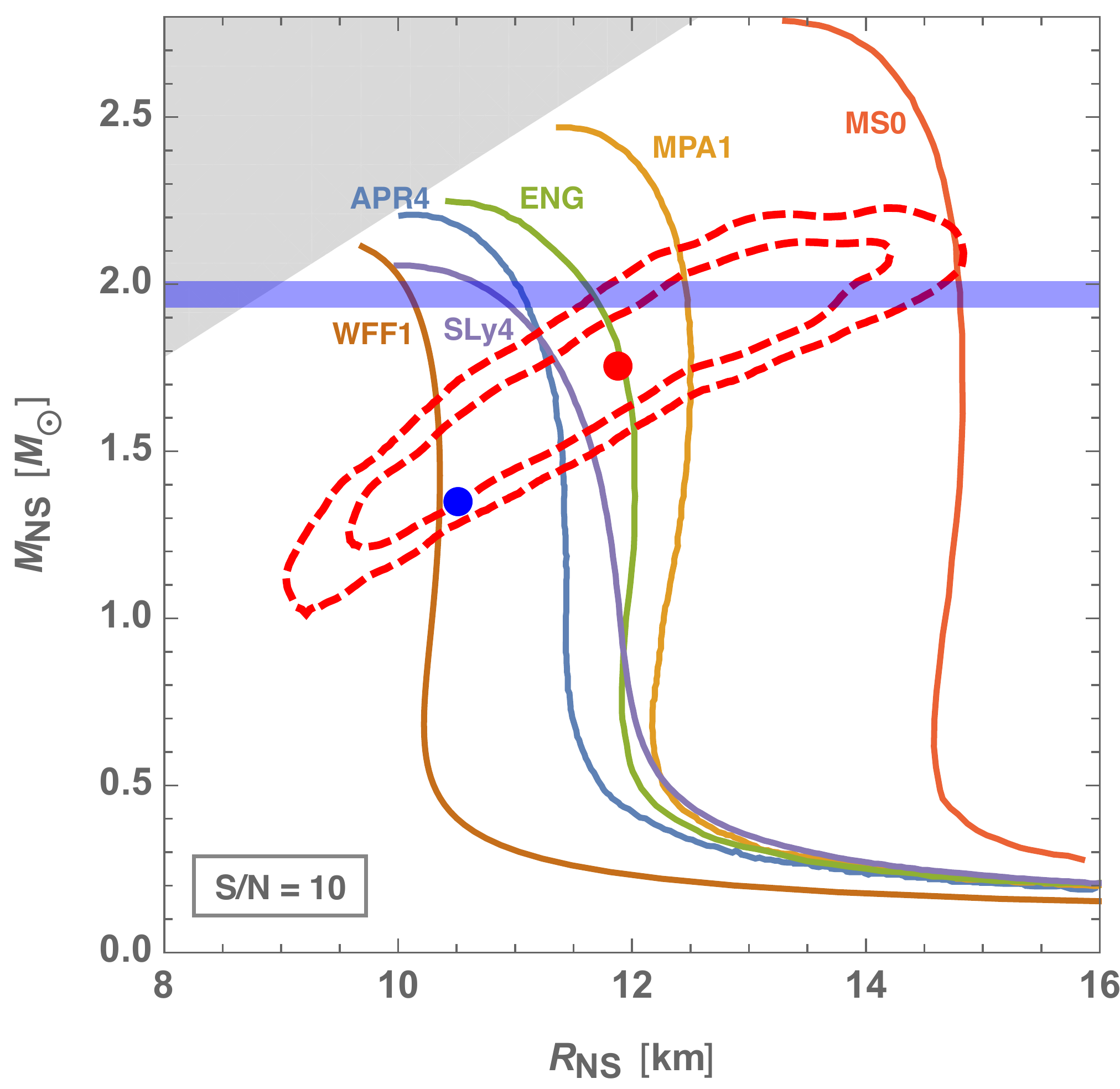}
  \caption{\ac{NS} mass and radius constraints obtained with our
    method for case \texttt{m484chi048H} and $\rm{\ac{SNR}}=30$ and
    $\rm{\ac{SNR}}=10$ in the top and bottom panel, respectively.
    \ac{NS} equilibrium sequences for different \ac{NS} \acp{EOS} are
    also shown. The shaded gray region discards mass-radius combinations excluded by the causality constrain. The horizontal blue band denotes the mass of the high-mass \ac{NS} J1614-2230~\citep{Demorest2010}. The dashed red lines represent the $68\%$ and $90\%$ credible
    regions. The blue dot marks the injected mass and radius values,
    while the red dot denotes the values recovered by the analysis as
    the median of the mass and radius distributions.}
  \label{fig:MvsRResult}
\end{figure}

Figure \ref{fig:HistoResult} shows the $\rNS$ posterior distribution
obtained for case \texttt{m484chi048H} ({\it i.e.},
$\mBH=4.84\,M_\odot$, $\mNS=1.35\,M_\odot$, $\spin=0.48$,
$\EISO=5\times10^{51}\,$erg): the top and bottom panel correspond to
$\rm{\ac{SNR}}=30$ and $\rm{\ac{SNR}}=10$, respectively.  The blue
solid line marks the target value of the radius, while the red solid
line marks $\lambda$, the median of the posterior.  Finally, the red
dashed lines mark the 5th and 95th percentiles of the posterior
distribution ($\lambda_{-}$, $\lambda_{+}$, with
$\lambda_{-}<\lambda_{+}$), which enclose the 90\% credible interval.
With this choice, the statistical error on the measurement is given by
\begin{equation}
  \sigma_{\rm Stat}\equiv\frac{\lambda_{+}-\lambda_{-}}{2\lambda}\,.
  \label{eq:RelError}
\end{equation}
We see that the 90\% credible interval encloses the target value of
$\rNS$ and that, as expected, it decreases as the \ac{SNR} increases.
Similarly, the difference between the injected value of $\rNS$ and the
median of the $\rNS$ posterior decreases with increasing \ac{SNR}.
These dependencies on \ac{SNR} are a sign of the impact that the our
\ac{GW}-informed prior for $\mNS$, $\mBH$, and $\spin$ has on the
final results of our approach.  We will return to this point in
Sec.\,\ref{sec:errors}.

In Fig.\,\ref{fig:MvsRResult}, the results for case
\texttt{m484chi048H} are displayed in the $\mNS$--$\rNS$ plane and
overlaid on \ac{NS} equilibrium sequences obtained with the APR4
\citep{Akmal1998a}, ENG \citep{ENG}, MPA1 \citep{MPA}, MS0 \citep{MS},
SLy4 \citep{SLy4}, and WFF1 \citep{WFF1} \ac{NS} \acp{EOS}.  Here the gray shaded area denotes the region of the $\mNS-\rNS$ plane where the causality constrain is not satisfied, while the blue horizontal band reports the mass of the millisecond pulsar J1614-2230, one of the \acp{NS} with the highest mass observed \citep{Demorest2010}. We can see that all the \acp{EOS} considered in this figure can account for this high value of mass. The red
dashed contours represent the $68\%$ and $90\%$ credible regions.  As
expected, this region shrinks as the \ac{SNR} increases, while still
including the injected values of mass and radius (blue dot).

\begin{figure}[!t]
  \centering
  \includegraphics[width=\columnwidth]{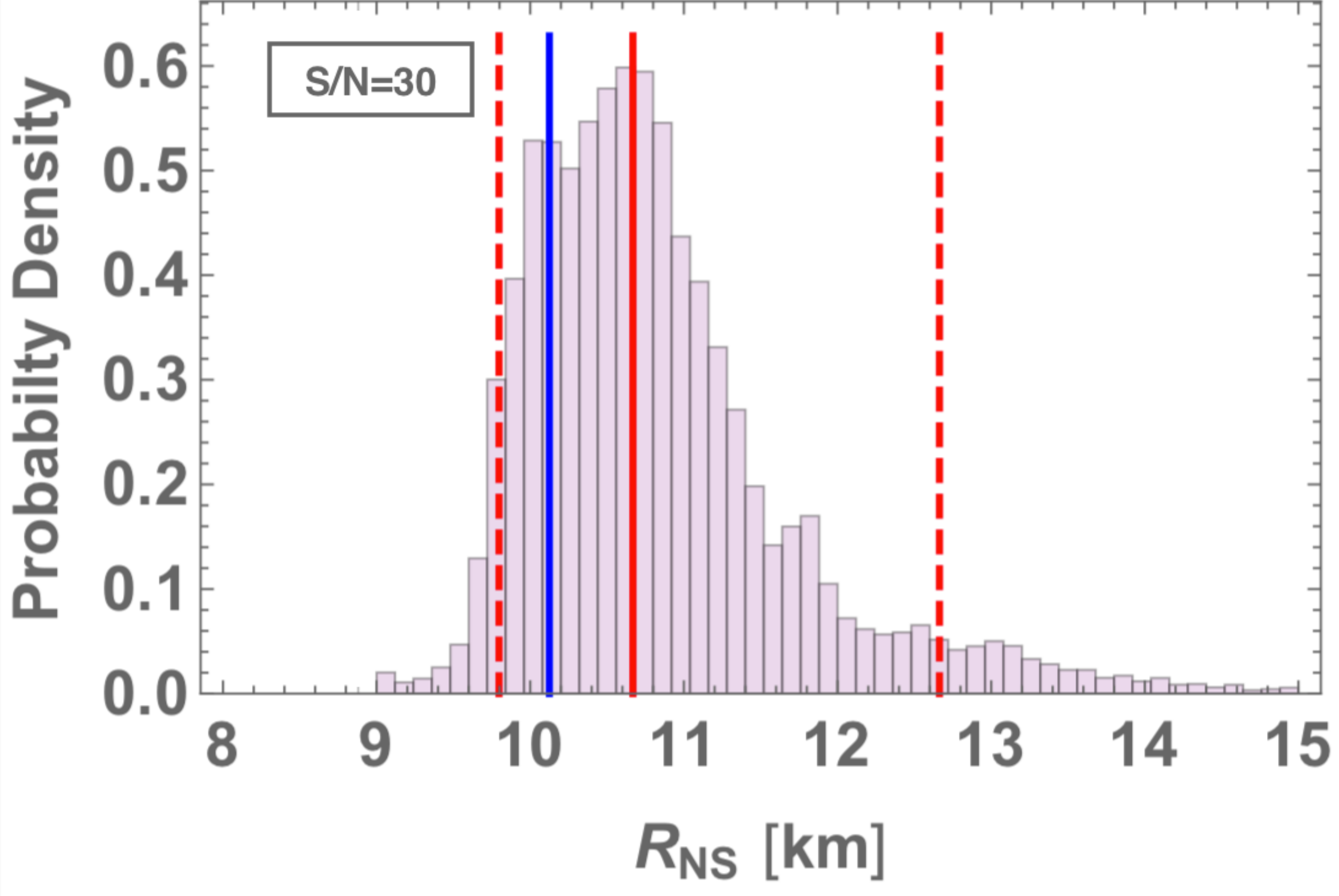}\\
  \includegraphics[width=\columnwidth]{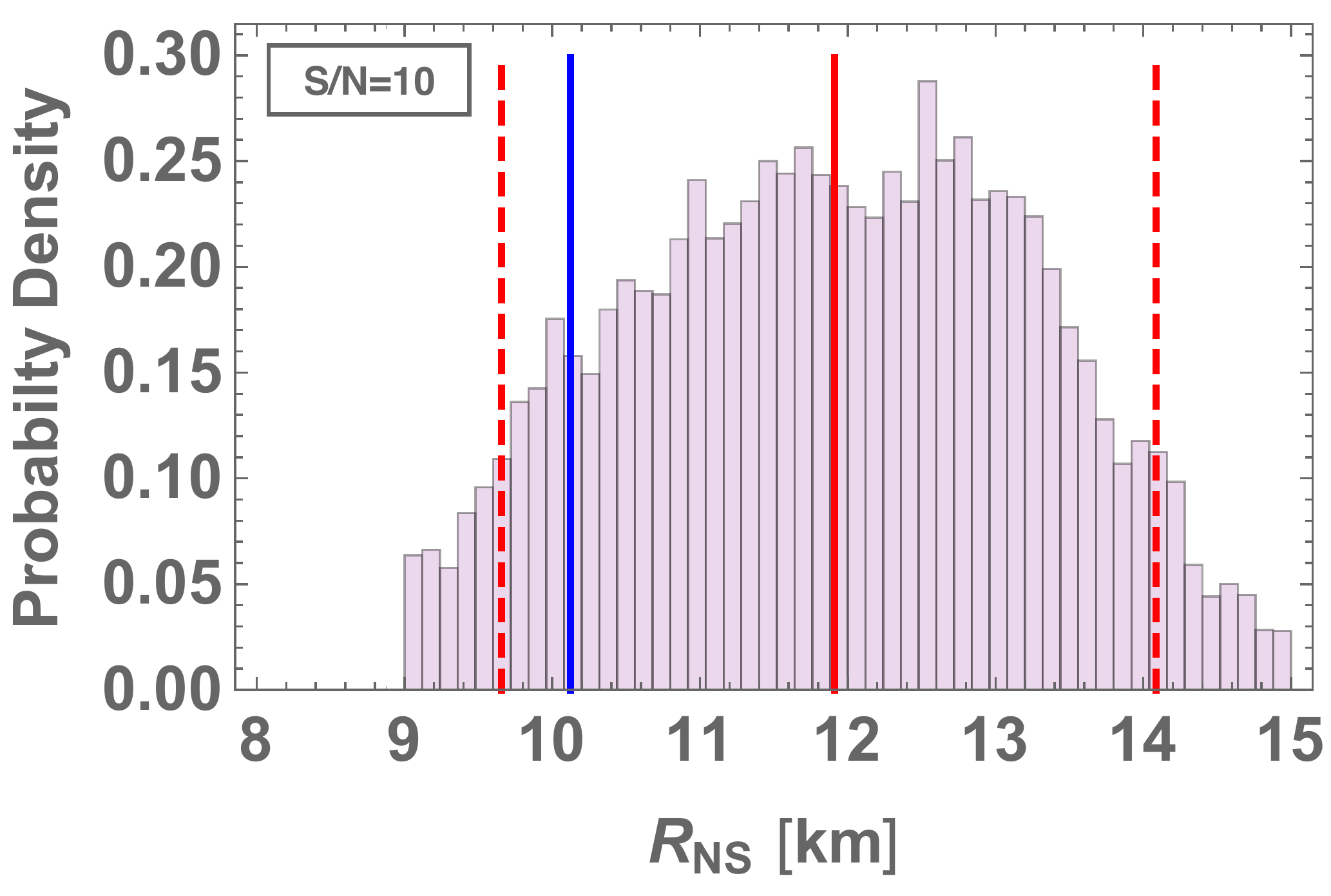}
  \caption{Same as Fig.\,\ref{fig:HistoResult} but for case
    \texttt{m484chi048L}.}
  \label{fig:HistoResult_51}
\end{figure}

\begin{figure}[!t]
  \centering
  \includegraphics[width=\columnwidth]{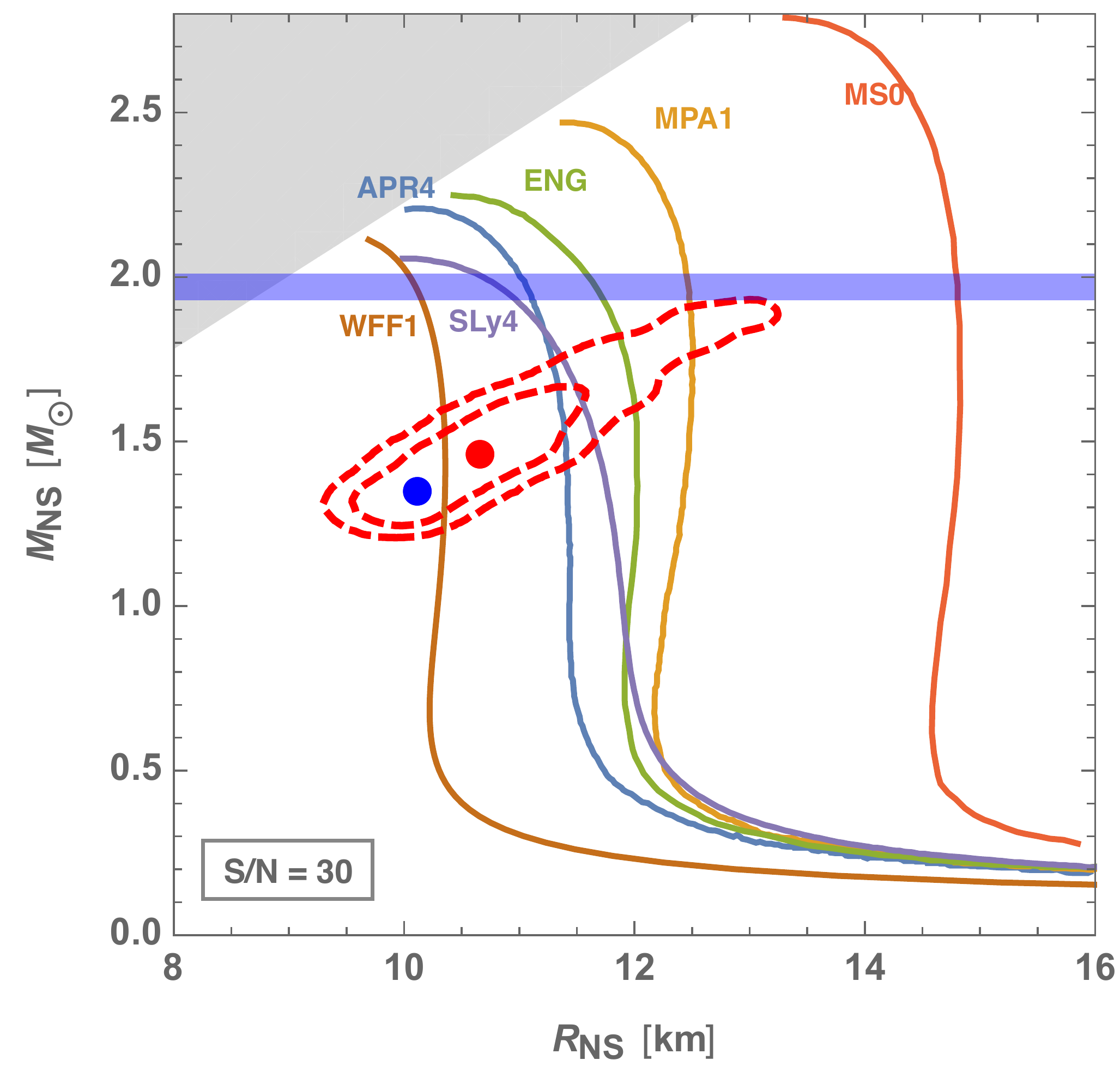}\\
  \vspace{0.2cm}
  \includegraphics[width=\columnwidth]{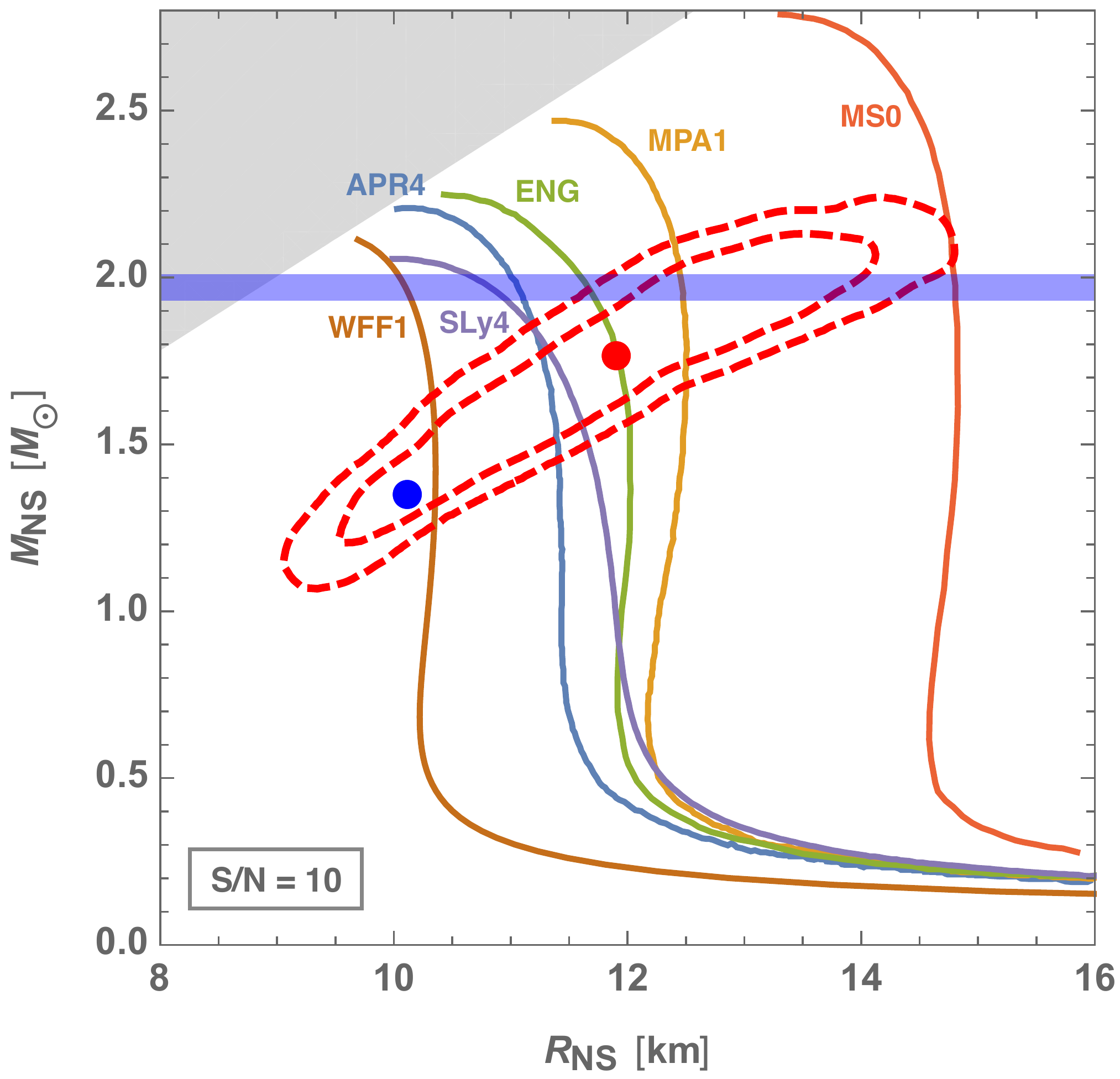}
  \caption{Same as Fig.\,\ref{fig:MvsRResult}, but for case
    \texttt{m484chi048L}.}
  \label{fig:MvsRResultH}
\end{figure}

Similar results hold for case \texttt{m484chi048L} and are shown in
Figures \ref{fig:HistoResult_51} and \ref{fig:MvsRResultH}. The
decrease in \ac{SGRB} energy causes the high-end tails of the $\rNS$
distribution to be slightly less populated with respect to the
\texttt{m484chi048H} case. This is not surprising: powering a more
energetic \ac{SGRB} requires a more massive torus, and lower values of
$\epsilon$ can accommodate larger values of $\rNS$ in such a
scenario. In turn, this means that the impact of the prior on
$\epsilon$ progressively increases with the \ac{SGRB} energy.

\begin{figure}[!t]
  \centering
  \includegraphics[width=\columnwidth]{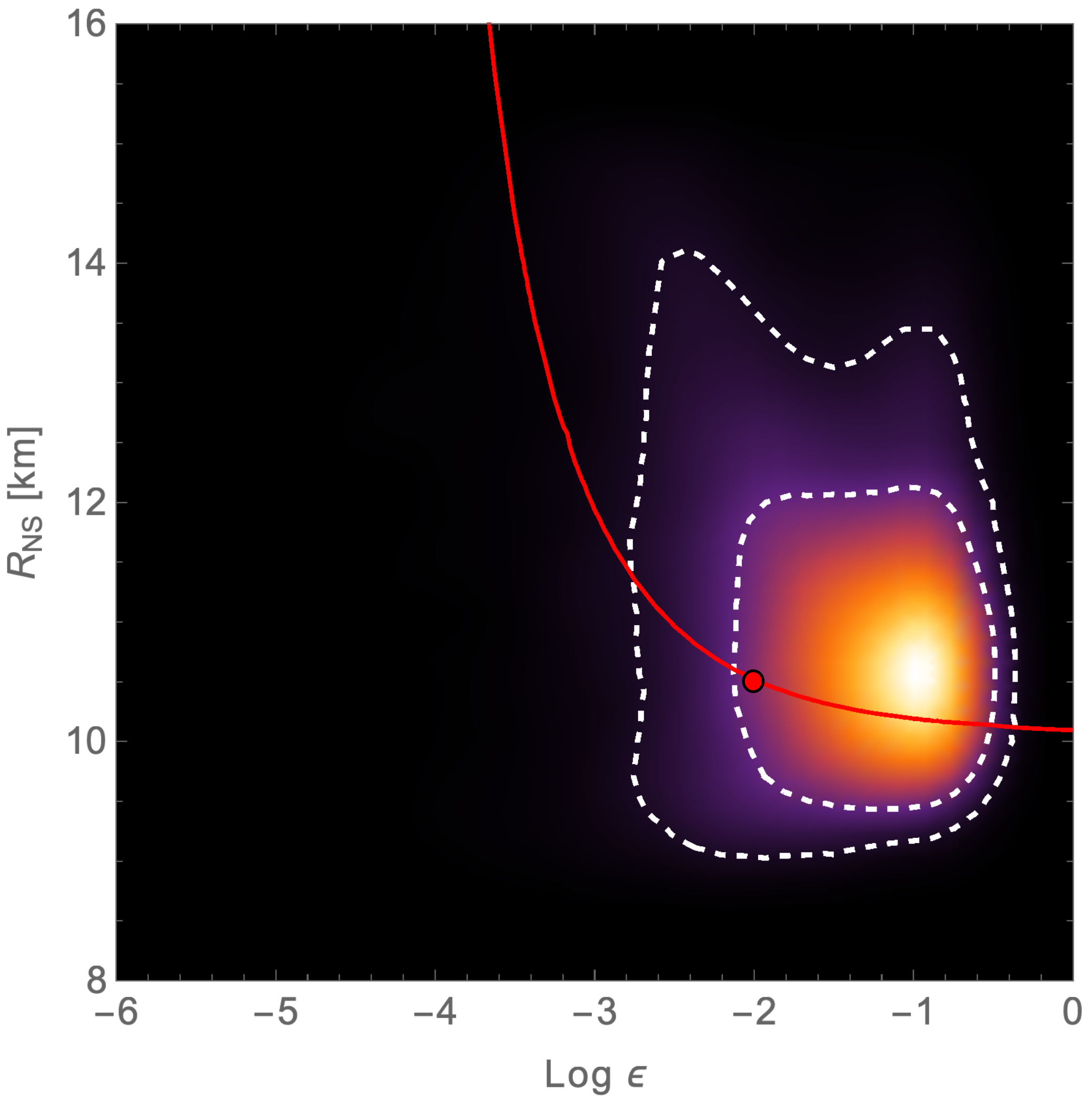}\\
  \includegraphics[width=\columnwidth]{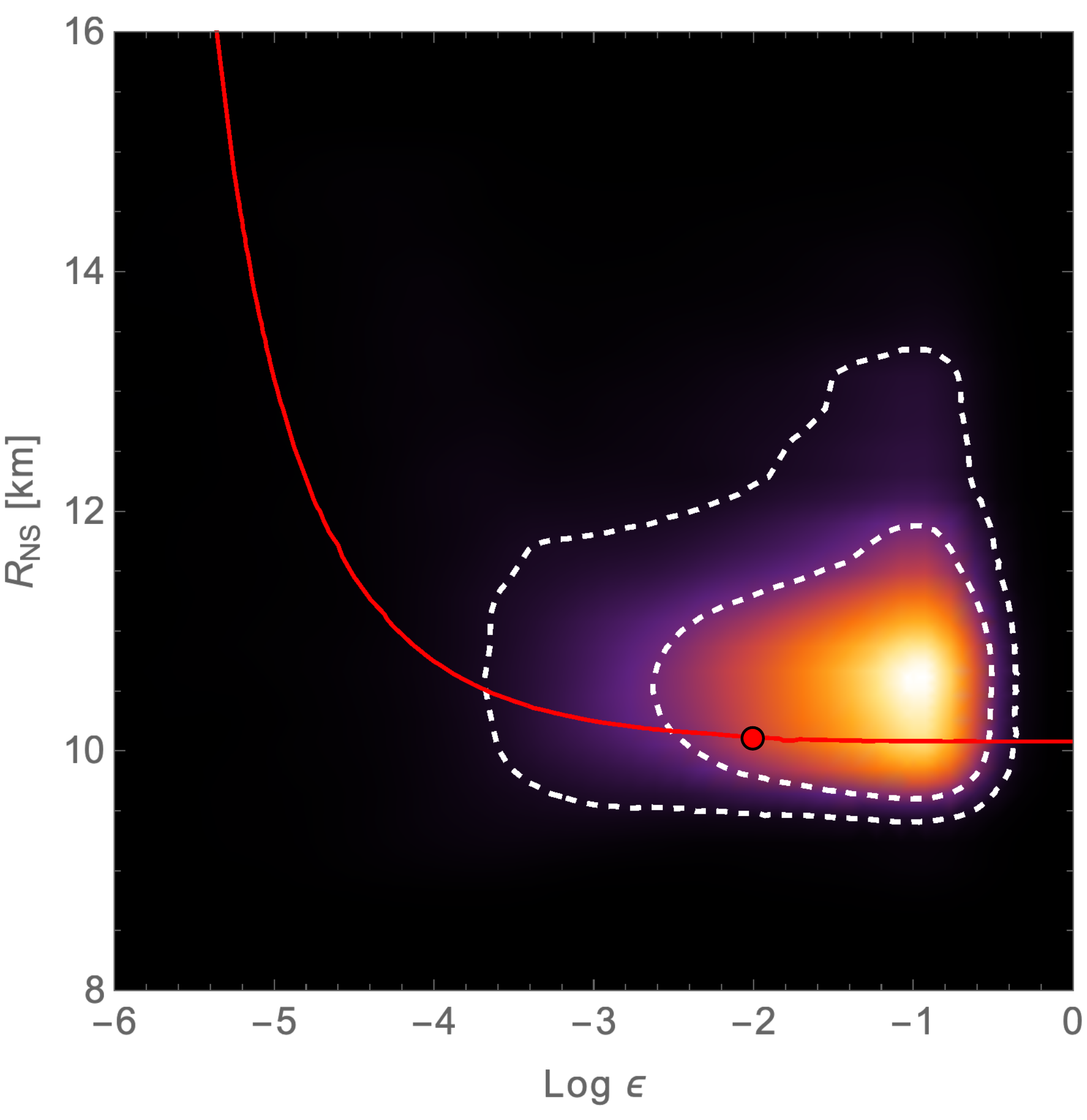}
  \caption{\ac{NS} radius and mass-energy conversion efficiency
    $\epsilon$ constraints for case \texttt{m484chi048H} (top panel)
    and case \texttt{m484chi048L} (bottom panel) with ${\rm
      \ac{SNR}}=30$. The white dashed curves represents the 68\% and
    90\% credible regions, respectively.  The red solid curves are the
    isoenergetic curve of the injection. The red dot marks the value
    of the injected epsilon and $\rNS$.}
  \label{fig:epsRad}
\end{figure}

This can be further understood from Fig.\,\ref{fig:epsRad}, where the
recovered posterior distributions for the high-energy case
\texttt{m484chi048H} (top panel) and the low-energy case
\texttt{m484chi048L} (bottom panel) are compared in the
$\epsilon$--$\rNS$ plane\footnote{We focus on this specific
  marginalization of the full results, because $\epsilon$ is the most
  influential among the unknown parameters that enter our method, and
  at the same time the least constrained by observations.} at ${\rm
  \ac{SNR}}=30$.  The red dot marks the simulated scenario, while the
white, dashed lines denote the 68\% and 90\% credible regions.  In the
low-energy case, the distribution is populated in regions with
$\epsilon \lesssim 10^{-3}$, so that the overall weight of high $\rNS$
values is reduced with respect to the high-energy case.  Furthermore, an
$\epsilon \lesssim 0.1\%$ gradually becomes unable to accommodate the
high-energy scenario, while this is not the case for the low-energy
case.  Finally, the red line is the curve of constant $E_\gamma$
(isoenergetic curve) obtained from Eq.\,(\ref{eq:reverseFouc}) for
this specific simulated scenario ({\it i.e.}, for $\mBH =
4.84\,M_\odot$, $\mNS=1.35\,M_\odot$, $\spin=0.48$, $\EISO =
10^{50}\,$erg, $\thj\simeq11^\circ$, $c_2 = 17/200$, $\epsilon=0.01$).
The fact that this curve cuts through the 68\% credible region shows
that our analysis is capable of recovering the simulated scenario.

\begin{figure}[!t]
  \centering
  \includegraphics[width=\columnwidth]{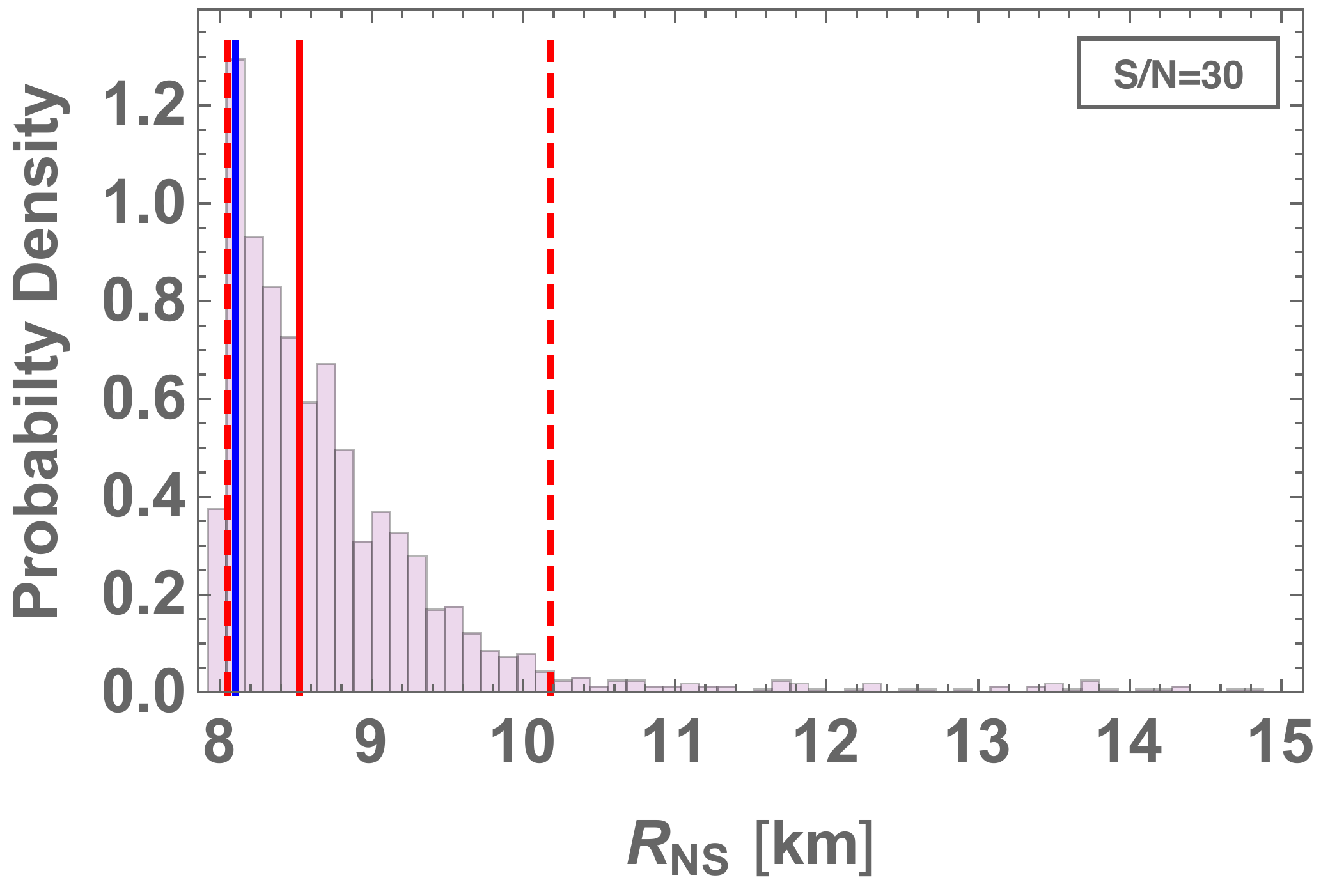}\\
  \includegraphics[width=\columnwidth]{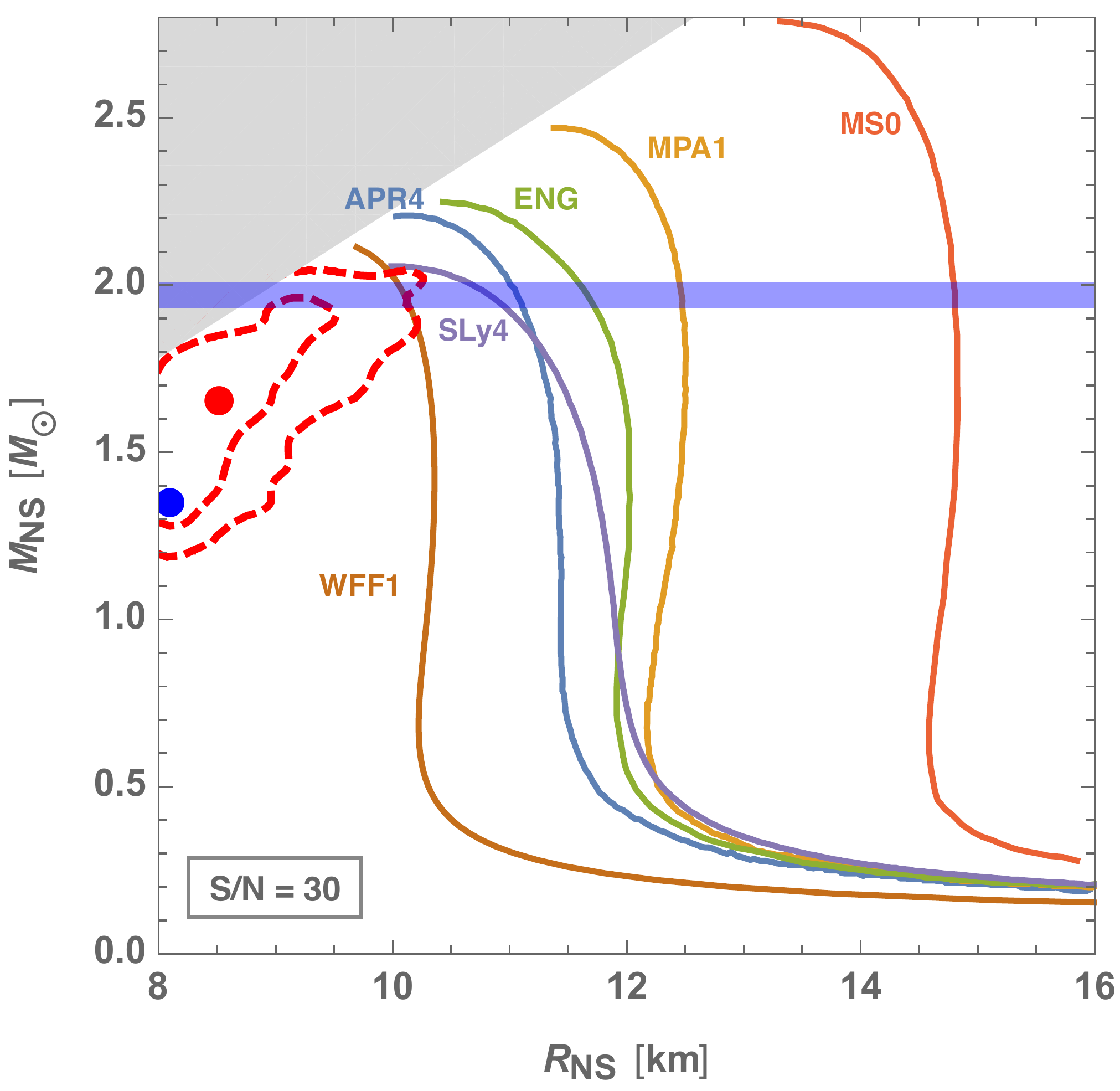}
  \caption{$\rNS$ posterior distribution (top panel) with the 68\% and
    90\% credible regions in the $\mNS$--$\rNS$ plane (bottom panel) for
    case \texttt{m484chi080H}.}
  \label{fig:HistoResult_5_0p8}
\end{figure}

We now vary the injected \ac{BH} parameters ($\mBH$ and $\spin$) to
see how this affects the recovery of $\rNS$.  We begin from the
\ac{BH} spin.  Figure \ref{fig:HistoResult_5_0p8} reports the results
at ${\rm\ac{SNR}}=30$ for case \texttt{m484chi080H}.  A comparison
with the \texttt{m484chi048H} results (Figs.\,\ref{fig:HistoResult}
and \ref{fig:MvsRResult}, top panels) highlights that, as the \ac{BH}
spin increases from $\spin=0.48$ to $\spin=0.8$, the $\rNS$ posterior
distribution shifts to lower values, correctly following the injected
$\rNS$ value\footnote{All else being fixed, an increase in $\spin$
  requires a decrease in $\rNS$ to maintain the \ac{SGRB} energy as
  constant.}.  In this particular case, where the value of the
injected $\rNS$ is small (see row 4 in Table \ref{tab:cases}), results
are obtained by extending the prior on $\rNS$ down to $8$\,km in order
to avoid a railing of the posterior distribution against the standard
boundary at $9\,$km.

\begin{figure}[!t]
  \centering
  \includegraphics[width=\columnwidth]{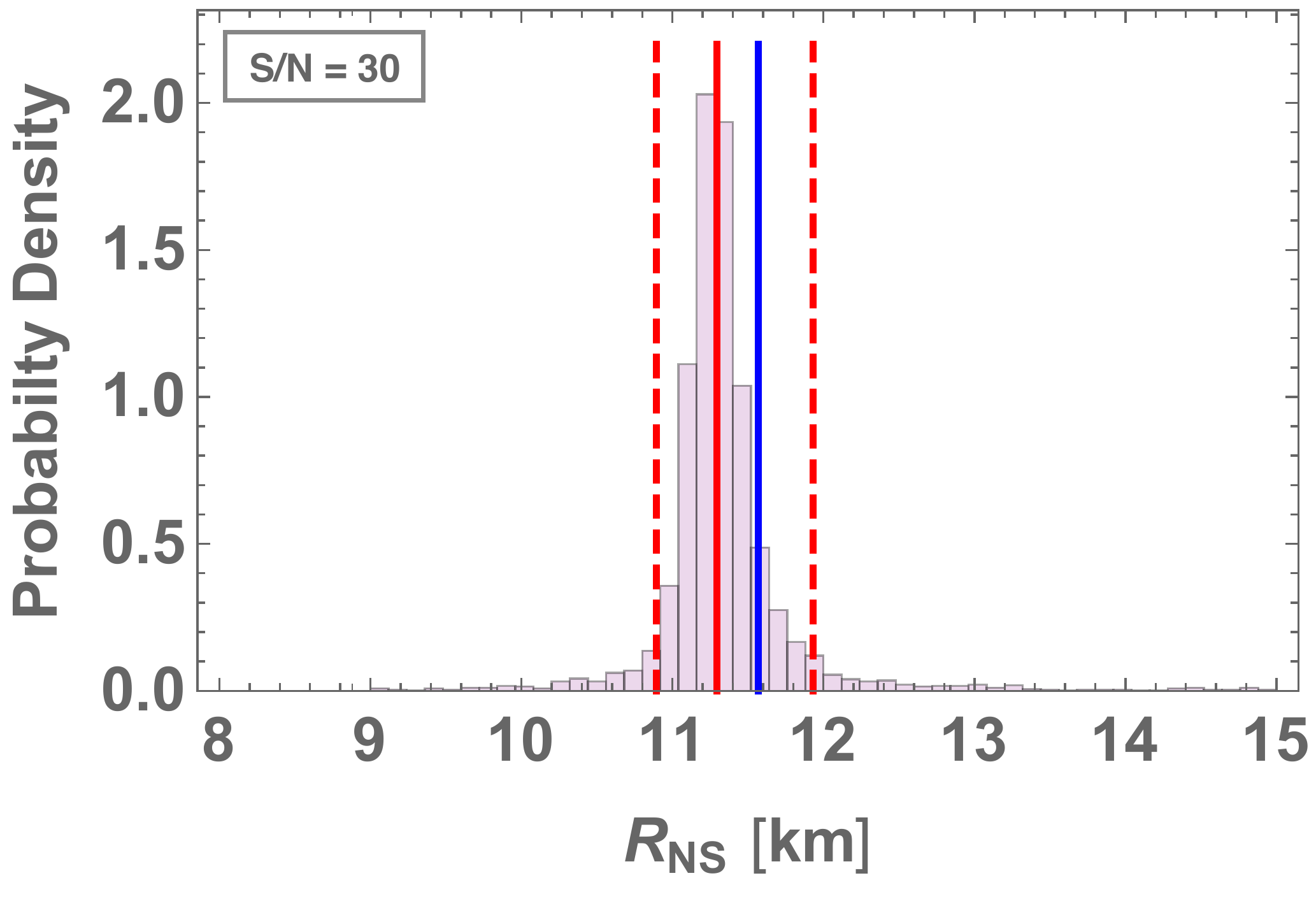}
  \includegraphics[width=\columnwidth]{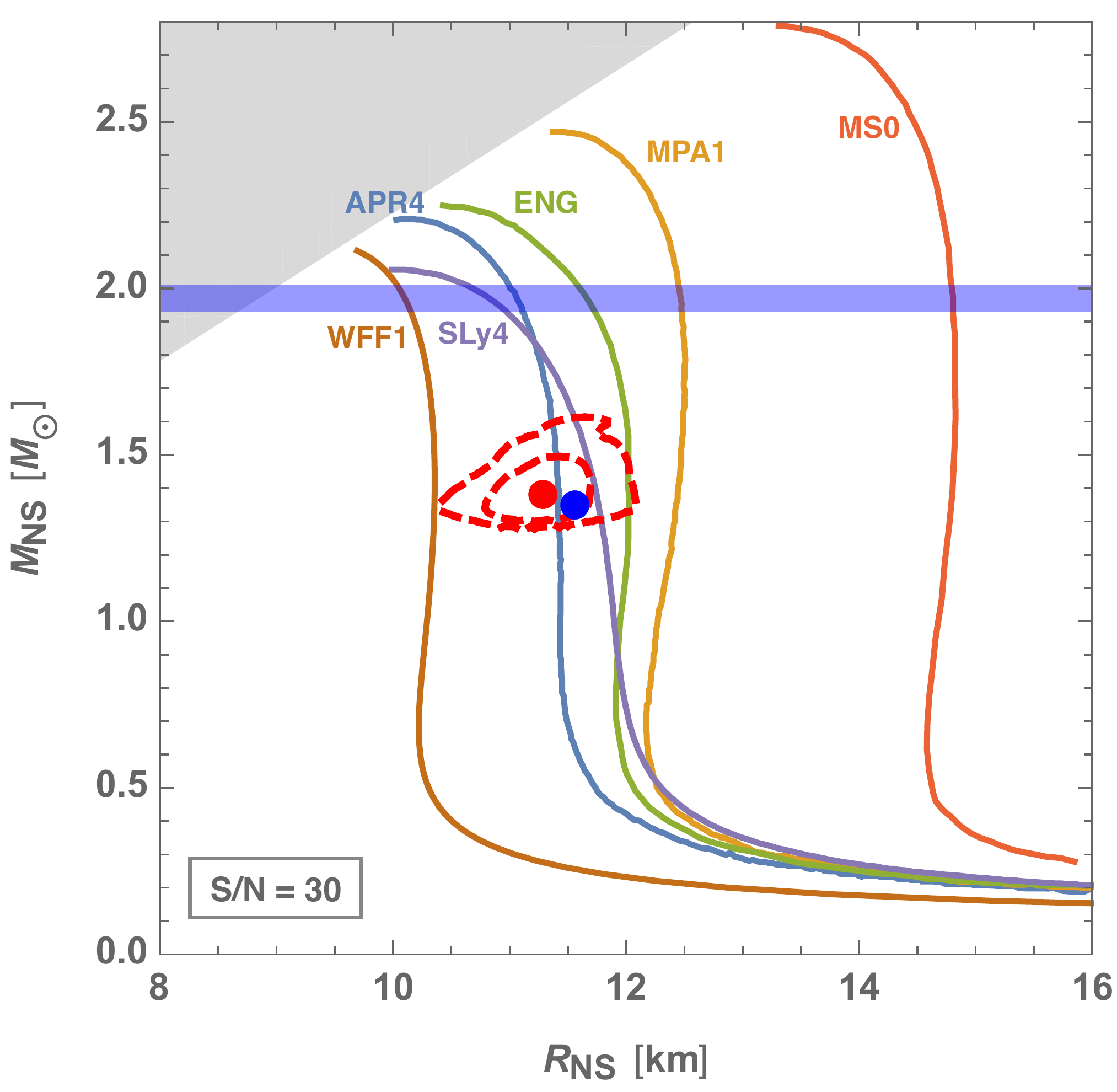}
  \caption{Same as Fig.\,\ref{fig:HistoResult_5_0p8} but for the
    \texttt{m100chi070H} case.}
  \label{fig:HistoResult_10_0p7}
\end{figure}

Figure \ref{fig:HistoResult_10_0p7} shows the results for case
\texttt{m100chi070H}, {\it i.e.}, the \ac{BH} has a rather high mass
and spin ($\mBH = 10\,M_\odot$, $\spin=0.7$).  In this case, the
\ac{BH} mass increase requires a higher simulated $\rNS$ value, and
the $\rNS$ posterior distribution accordingly shifts toward higher
values.

\subsection{Accuracy of the $\rNS$ measurement} \label{sec:errors}

In this section, we address the impact of the \ac{GW} posterior, which
we use as an informed prior for our method, on the measurement of
$\rNS$.  Furthermore, we discuss the overall uncertainty on the \ac{NS}
radius recovered with our approach.

\begin{figure}[!h]
  \centering
  \includegraphics[width=\columnwidth]{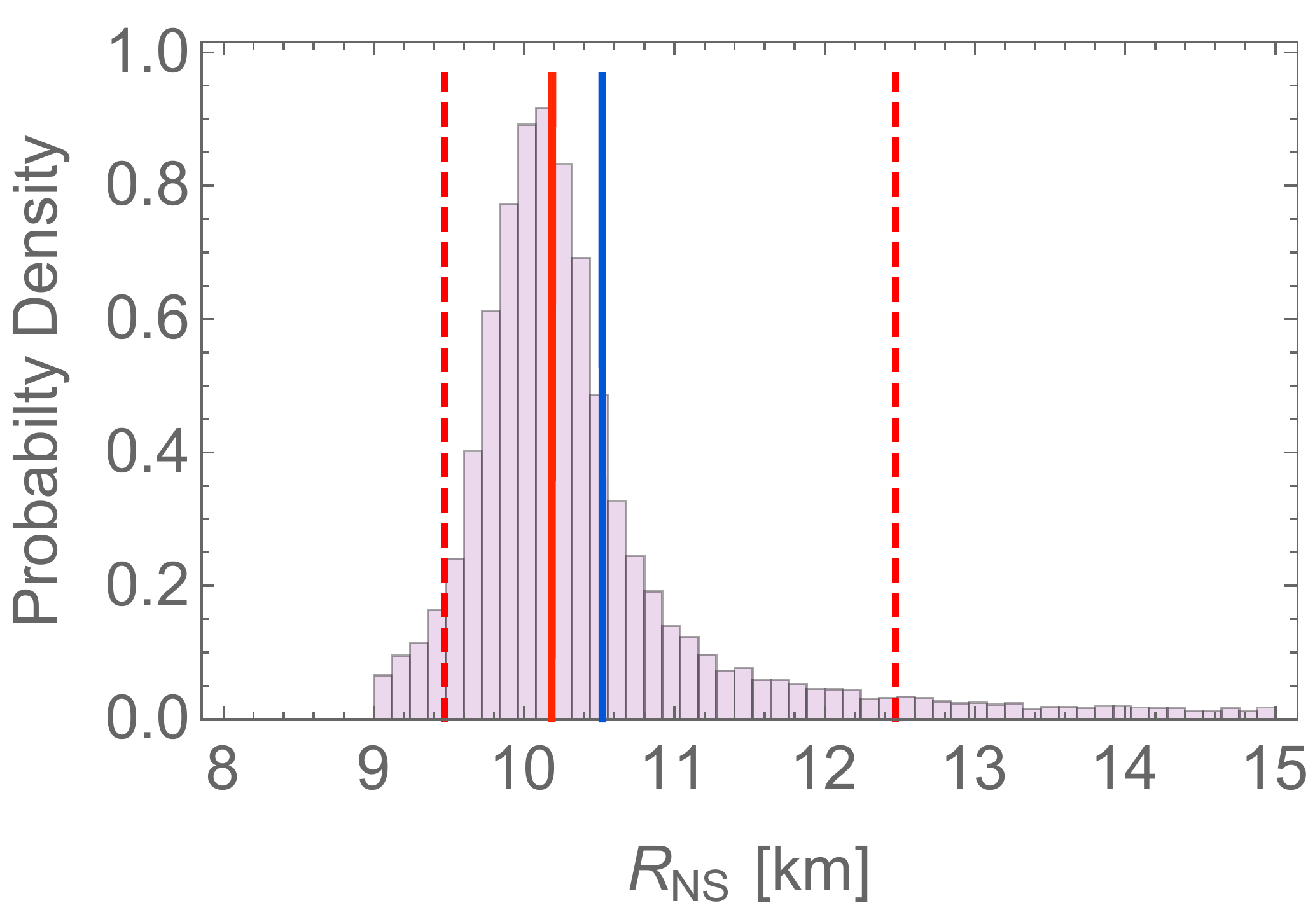}\\
  \includegraphics[width=\columnwidth]{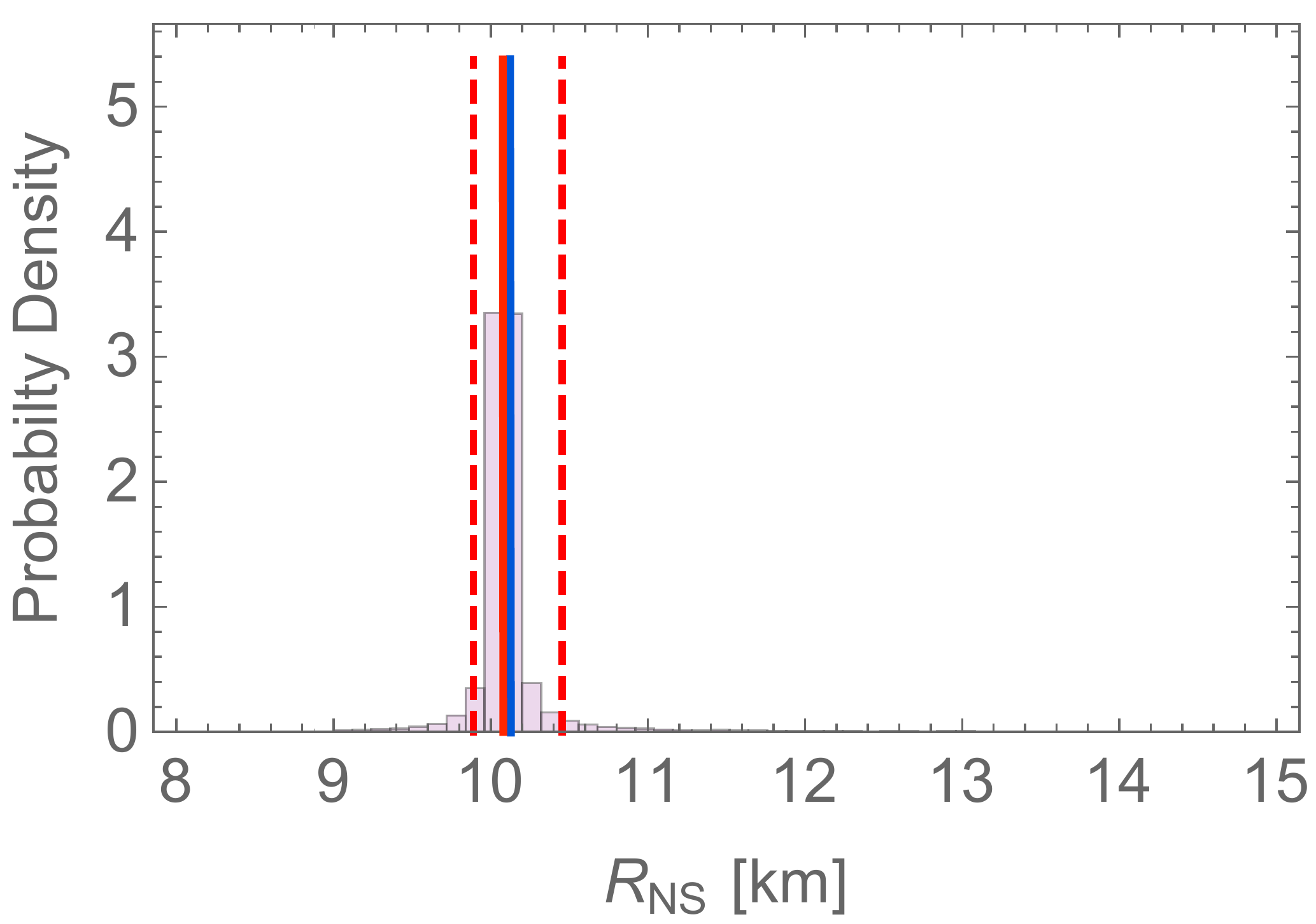}
  \caption{$\rNS$ posterior obtained when assuming $\mNS$, $\mBH$, and
    $\spin$ to be known exactly for cases \texttt{m484chi048H} ({\it
      top}) and \texttt{m484chi048L} ({\it bottom}).}
  \label{fig:rNSandPriors}
\end{figure}

Figure \ref{fig:rNSandPriors} shows cases \texttt{m484chi048H} and
\texttt{m484chi048L} analyzed in the hypothetical scenario in which
$\mBH$, $\mNS$, and $\spin$ are known exactly (which makes the
\ac{SNR} value irrelevant).  In other words, we set to zero any
systematics deriving from the \ac{GW} informed prior, but we sample
$\epsilon$, $\thj$, and $c_2$ normally.  This allows us to quantify how
the analysis of the \ac{GW} data influences our final result.  The
upper and bottom panel of this figure should be compared to the panels
in Figs.\,\ref{fig:HistoResult} and \ref{fig:HistoResult_51},
respectively. In the high-energy case, the recovered median now
slightly underestimates the injected value of $\rNS$, and the width of
the posterior is reduced.  The change in width of the posterior is
even more dramatic for the low-energy case, which now displays a
virtually perfect recovery of the injected value.

\begin{figure}[!h]
  \centering
  \includegraphics[width=\columnwidth]{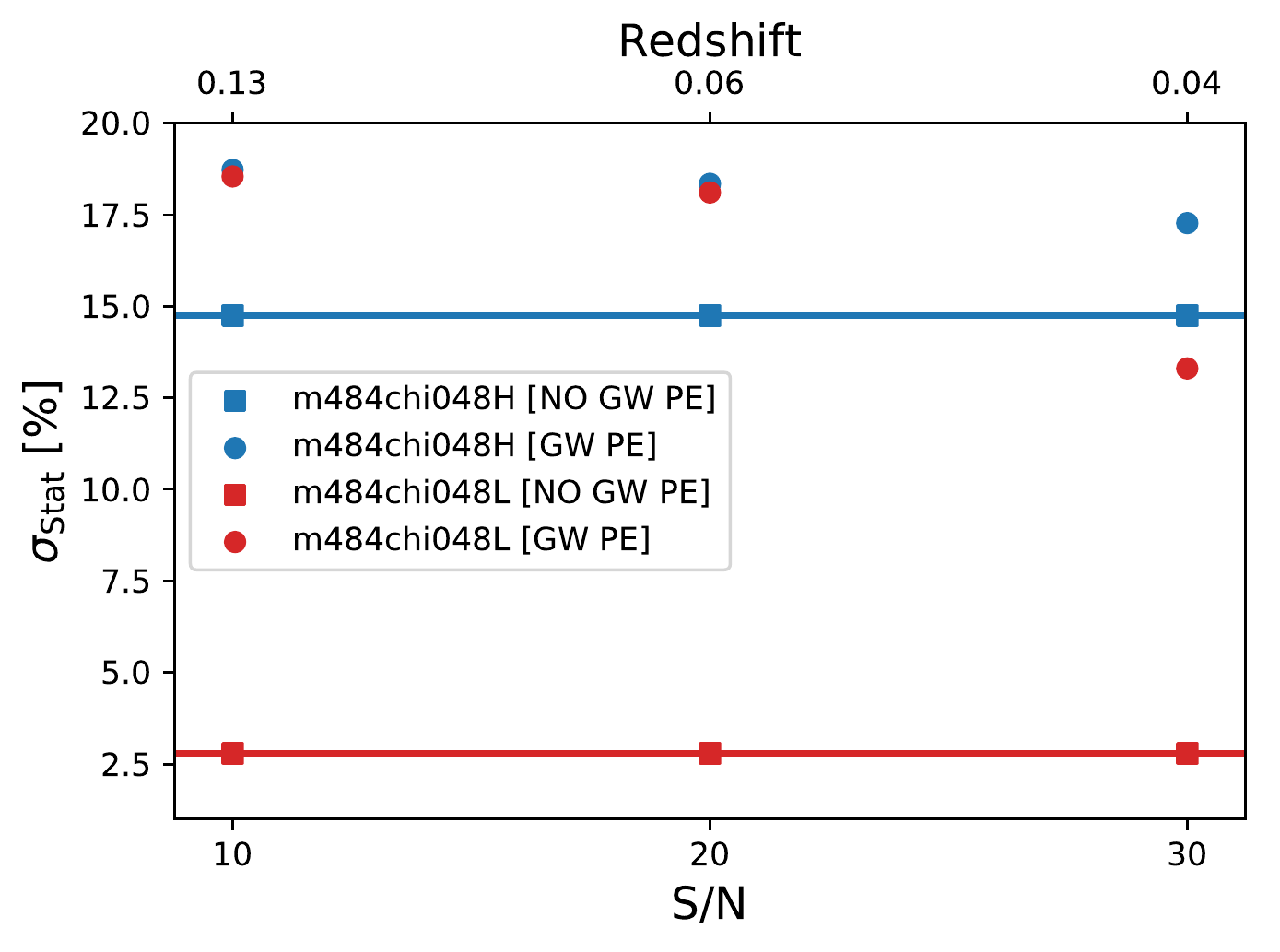}
  \includegraphics[width=\columnwidth]{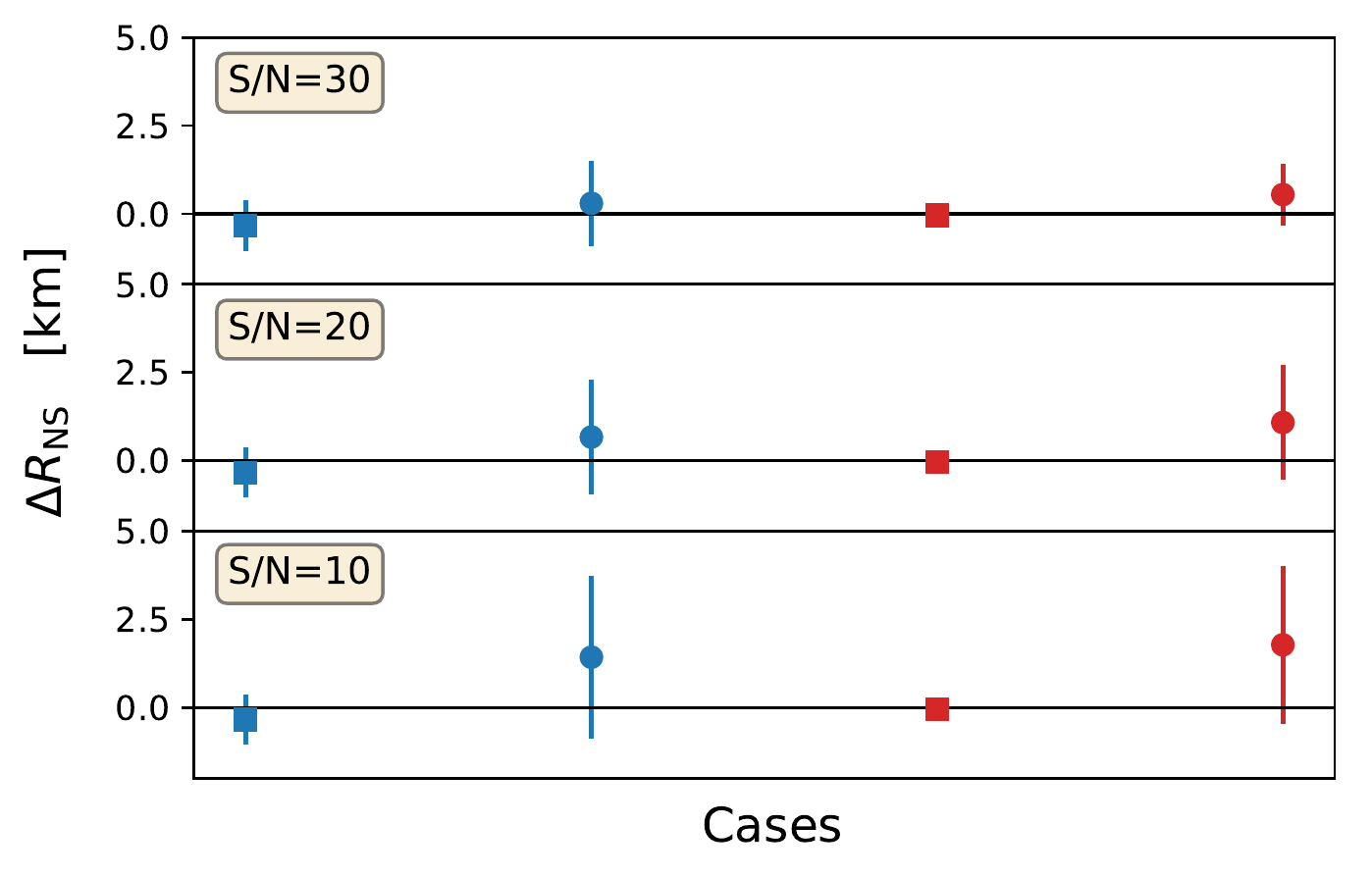}
  \caption{{\it Top panel:} statistical error
    (Eq.\,(\ref{eq:RelError})) on $\rNS$ as function of the \ac{GW}
    \ac{SNR}. The blue and red markers denote cases \texttt{m484chi048H}
    and \texttt{m484chi048L}, respectively. The circles (squares)
    represent cases that use (do not use) the prior on $\mNS$, $\mBH$,
    and $\spin$ informed by \ac{GW} parameter estimation [which is
    denoted as ``GW PE'' in the legend]. {\it Bottom panel}: error on
    $\rNS$ for all the scenarios considered in the top panel; the
    symbols denote the systematic error, that is, the difference
    between the median and the injected value, while the bars indicate
    the 90\% credible intervals, {\it i.e.}, the statistical
    uncertainty.}
  \label{fig:RelError}
\end{figure}

The top and bottom panels in Fig.\,\ref{fig:RelError} show how the
statistical error on $\rNS$, as defined in Eq.\,(\ref{eq:RelError}),
and the systematic error on $\rNS$ vary with the \ac{SNR} of the
\ac{GW} signal for cases \texttt{m484chi048H} (blue) and
\texttt{m484chi048L} (red), when using the \ac{GW} informed prior
(circles) and when, instead, assuming that the two masses and the
\ac{BH} spin are known exactly (squares).  The statistical error on
$\rNS$ (top panel) for the \texttt{m484chi048H} and
\texttt{m484chi048L} standard analysis setup is well behaved as it
decreases with \ac{SNR}.  When we assume $\mNS$, $\mBH$, and $\spin$
to be known exactly, it clearly does not depend on the \ac{GW}
\ac{SNR}, hence the use of a continuous line at a constant value.  The
statistical error in the low-energy case is systematically lower than that
in the high-energy case. As discussed in Sec.\,\ref{sec:epsilonPrior},
this happens because the results for the low \ac{SGRB} energy case
depend more weakly on $\epsilon$.  Since $\thj$ and $c_2$ enter
Eq.\,(\ref{eq:reverseFouc}) in the same term as $\epsilon$, the same
argument may be applied to these two parameters.  Overall, at lower
\ac{SGRB} energy the impact of $\epsilon$, $\thj$, and $c_2$ on the
final result is weaker, which in turn means that the statistical
uncertainty on $\rNS$ is expected to decrease.  As demonstrated in the
top panel of Fig.\,\ref{fig:RelError}, this also implies that within
our approach the \ac{SGRB} energy determines a lower bound on the
statistical error on $\rNS$ that cannot be beaten by increasing the
\ac{GW} \ac{SNR}.  Furthermore, this bound decreases with the \ac{SGRB}
energy.  Therefore, for low energies the uncertainties on $\mNS$,
$\mBH$, and $\spin$, which derive solely from the analysis of the
\ac{GW} data, end up dominating the accuracy of the measurement of
$\rNS$.  The bottom panel shows that, unsurprisingly, the bias in the
measurement of $\rNS$ is larger when using the \ac{GW} informed prior,
as opposed to when $\mNS$, $\mBH$, and $\spin$ are assumed to be known
exactly.  As expected, the overall bias decreases with \ac{SNR}.
Finally, by contrasting results for which we assume to know the values
of $\mNS$, $\mBH$, and $\spin$ (squares) to results that are not based
on this assumption (circles), we see that the bias introduced by the
\ac{GW} analysis acts in the direction opposite of that of the bias
introduced by the second step of our hierarchical method, {\it i.e.},
sampling of $\epsilon$, $\thj$ and $c_2$ and use of
Eq.\,(\ref{eq:Foucart}).

Our lack of knowledge about $\thj$ and $\epsilon$ contributes in
shaping the $\rNS$ posterior distribution.  Therefore, in the event of
a joint \joint \nsbh observation, any input from additional \ac{EM}
observations and from theoretical studies about jet-launching
mechanisms could lead to improvements in the $\rNS$ posterior
distribution. Similarly, detailed analyses of the \ac{GW} alone could
also improve the radius measurement further by providing a tighter
informed prior for $\rNS$~\citep{GW170817EOS, GW170817PE}.

Finally, we wish to stress that, unfortunately, a proper assessment of
all the systematics that enter our method is currently unfeasible.
A first assessment of systematics could be achieved as follows.  One
would have to run numerical-relativity simulations of various \nsbh
mergers, extract the remnant masses from them, build complete \ac{GW}
signals by combining analytic approaches for the early inspiral with
the numerical data for the late inspiral and merger, and finally test
our method against such signals and remnant mass values\footnote{All
  this would be done by fixing the value of $\epsilon$ in order to
  determine the \ac{SGRB} energy, as no simulation from the initial
  \nsbh binary to the final \ac{SGRB} is currently possible.}.  This
extensive investigation is beyond the scope of the present work and we
leave it as a topic for future studies.  Because it would heavily rely
on numerical-relativity simulations, this would only be a first,
albeit significant step.  Importantly, in this context,
\citet{Foucart2018} found no systematic bias associated with the
numerical-relativity code used to determine remnant mass values and
that different codes predict remnant masses to within the accuracy of
Eq.\,(\ref{eq:Foucart}).

\section{Discussion}
\label{sec:discussion}

The joint observation of GW170817 and GRB 170817A has unambiguously
associated \bns coalescences and \acp{SGRB}~\citep{GW170817GRB}
confirming the long-standing hypothesis that \bns binaries are
\ac{SGRB} progenitors~\citep{Blinnikov1984, ELPS89, Paczynski1986,
  ELPS89, P91, NPP92}.  While the rate of \bns mergers can accommodate
for the rate of observed \ac{SGRB} events~\citep{GW170817GRB}, the
question of whether \acp{SGRB} have more than one kind of progenitor
remains an open one, and one that future observing runs of current and
upcoming \ac{GW} detection facilities will help answer.
\nsbh systems, in particular, remain a viable \ac{SGRB} progenitor
candidate (see, {\it e.g.}, \citet{Nakar:2007yr}).
\citet{C15} determine a projected joint \joint detection rate for
\nsbh coalescences of $0.1$--$2\,$yr$^{-1}$ for Advanced LIGO and
Virgo at design sensitivity and the {\it Fermi} Gamma-Ray Burst
Monitor, which decreases to $0.03$--$0.7\,\mathrm{yr}^{-1}$ with
\emph{Swift}.  Similarly, \citet{Regimbau:2014nxa} found a joint \joint
detection rate with \emph{Swift} of $0.05$--$0.06\,\mathrm{yr^{-1}}$
while \citet{Wanderman:2014eza} found $0.4$--$1\,\mathrm{yr^{-1}}$
($3$--$6\,\mathrm{yr^{-1}}$ with {\it Fermi} Gamma-Ray Burst Monitor).
The next generation of \ac{GW} interferometers will extend the \nsbh
detection horizon up to $z\simeq4$~\citep{ETdesignstudy} therefore
boosting such detection rates.

In this paper, we presented a method based on
\citet{Pannarale:2014rea} to exploit joint \joint observations of
\nsbh coalescences in order to measure the \ac{NS} radius, and hence,
constrain the \ac{EOS} of matter at supranuclear densities. We sample
the \ac{GW} posterior distribution of the component masses and the
\ac{BH} spin along with uniform prior distributions on other unknown
physical parameters involved in the problem --- among which is the
\ac{NS} radius (see Sec.\,\ref{sec:methodology} for details) --- and
determine a distribution of isotropic gamma-ray energies.  This is
then combined with the \ac{EM} measurement of the isotropic gamma-ray
energy to yield a constraint on the \ac{NS} radius, after
marginalizing over all other sampled quantities.  \citet{Hinderer2018}
performed a similar analysis on GW170817, also using
\citet{Foucart2018} and working under the assumption that the event
originated from a \nsbh coalescence, but exploiting the \ac{EM}
constraints from the kilonova light curve, rather than the \ac{SGRB}
energy.

In order to test the performance and the robustness of our method, we
simulated six joint \joint \nsbh detection scenarios (see Table
\ref{tab:cases}). In each case, we compared the injected $\rNS$ value
to the posterior distribution recovered by our analysis.  While this
setup does not allow us to assess systematics in our methodology (see the
discussion at the end of Sec.\,\ref{sec:errors}), it is currently the
only possible benchmark and it allows us to draw the following first, important
conclusions about our method:
\begin{itemize}
\item The 90\% credible regions we determine always contains the
  injected value of $\rNS$, regardless of the mass and/or spin of the
  \ac{BH} in the \nsbh system under consideration.
\item With the exception of case \texttt{m100chi070H}, the median of
  the $\rNS$ posterior distribution is usually higher than the
  injected \ac{NS} value and it is narrower for lower-energy
  \acp{SGRB} ({\it i.e.}, $\EISO\lesssim 10^{50}\,$erg).
\item We can constrain the \ac{NS} radius with an uncertainty
  (quantified from a 90\% of credible interval) below 20\% even for
  low \ac{SNR} events.
\item The $\rNS$ lower bound is rather solid and depends mostly on the
  \ac{SNR} of the \ac{GW} signal through the informed prior for the
  \ac{GW} parameters.
\item By directly sampling the posterior distributions of \ac{GW}
  parameter estimation analyses, our method inherits any uncertainty
  that is present in such distributions.  This component of the
  overall error on the recovered $\rNS$ reduces as the \ac{SNR} of the
  \ac{GW} increases.  However, in Sec.\,\ref{sec:errors} we showed
  that the \ac{SGRB} energy determines a hard lower limit for the
  uncertainty on $\rNS$.  The value of this contribution to the
  overall error is clearly \ac{SNR} independent, but it decreases with
  the \ac{SGRB} energy.  For example, for the source configuration
  considered in Fig.\,\ref{fig:RelError}, this lower limit varies from
  $\sim 3\%$ to $\sim15\%$ as $\EISO$ goes from $10^{50} \, \rm erg/s$
  to $5\times 10^{51} \, \rm erg/s$.
\end{itemize}

A central ingredient of our method is the fitting formula that
predicts the mass of the matter that remains in the surroundings of
the remnant \ac{BH} immediately after the merger as a function of the
\nsbh initial parameters~\citep{Foucart2018}.  This can be replaced as
improved or different versions of such formula are published.
However, as long as it remains the only available option in the
literature, a study of systematics continues to be a time and resource
intensive task that would essentially require a campaign of
numerical-relativity simulations (see discussion at the end of
Sec.\,\ref{sec:errors}).  Furthermore, for such a study to be fully
self-consistent, one would require simulations that evolve the \nsbh
system all the way from inspiral to the ignition of the \ac{SGRB}.
For the time being, the tolerance we introduce in
Eq.\,(\ref{eq:tolerance}) when comparing our inferred $\EISO$ values
to the observed $\EISO$ accounts for systematic uncertainties in the
fit of \citet{Foucart2018}, but also for possible differences between
the remnant mass that it models and the disk mass that actually
accretes onto the central \ac{BH}. These two quantities may differ,
for instance, if a non-negligible fraction of remnant mass were to be
lost in form of dynamical ejecta or disk
winds~\citep{Kawaguchi:2016ana}.  Although our method is therefore
model-dependent, we note that this is a shared feature of all other
existing methods to measure \ac{NS} radii (for a recent review, see
\citet{OF16}).  For example, $\rNS$ constraints from low-mass X-ray
binary observations that are based on spectroscopic measurements of
such sources in a quiescent state~\citep{HRNG06, WB07, GRB11, BHOG16} or
after a thermonuclear burst~\citep{V79, OGP09, GWCO10, GOCW10, OGG12,
  GO13} require, among other things, introducing assumptions about the
\ac{NS} atmosphere composition and magnetic field.  Other methods that
involve timing measurements of oscillations in accretion-powered
pulsars
\citep{PG03, LMC08, LMCC09, LMC11, ML11} require modeling
the pulsed waveform and therefore depend on assumptions about \ac{NS}
spacetimes and other geometrical factors, such as the shape and
location of the surface hotspots.
Finally, \ac{EOS} constraints that rely on the analysis of \ac{GW}
data, including our method, intrinsically depend on the waveform
models used to process the \ac{GW} data and on how these treat tidal
effects~\citep{GW170817EOS, GW170817PE}.  These examples illustrate
that a model dependency is unavoidable when addressing the task of
measuring \ac{NS} radii.  However, the availability of a number of
methods each one of which relies on different assumptions and on the
observation of different astrophysical systems is crucial: the
combination of results that stem from various approaches can provide a
more solid, final result.

On the basis of the work carried out in this paper, there are a number
of lines of investigation that we plan to explore.  Firstly, in the
event of an \nsbh detection, a detailed analysis of the \ac{GW} that
constrains the \ac{NS} tidal deformability would be carried out, as
was the case for the \bns coalescence event GW170817~\citep{GW170817,
  GW170817EOS, GW170817PE}.  In turn, this information and the
so-called ``universal relations'' (see, {\it e.g.}, \cite{YY17}{ for a
  review}) could be exploited to build a less agnostic sampling of the
\ac{NS} radius to be used within our approach (currently a uniform
prior between $9\,$ and $15\,$km): upper limits on the tidal
deformability would result in a narrower interval to be sampled.
Moreover, this informed prior on $\rNS$ would also ensure a more
consistent sampling of the \ac{NS} mass and radius, with more massive
objects associated with higher compactnesses.  Furthermore, in the event
of an \nsbh merger observation in which the \ac{NS} is disrupted by
the \ac{BH} tidal field, the \ac{GW} signal is expected to shut off at
a characteristic frequency which depends, among other things, on the
\ac{NS} \ac{EOS}~\citep{Shibata:2009cn, KOST11, Pannarale2015a}.  The
measurement of this frequency would yield constraints on $\rNS$ with a
$10$--$40$\% accuracy~\citep{Lackey2014, LK14}, and we want to assess
the impact of including such information into our analysis.  This
scenario is particularly relevant for third-generation \ac{GW}
detectors because the shutoff of \nsbh signals happens in the
$\sim$kHz \ac{GW} frequency regime.  The projected \nsbh detection
rate for the Einstein Telescope is
$\mathcal{O}(10^3$--$10^7\,$yr$^{-1})$~\citep{ETdesignstudy}.  In
order to guarantee a high-joint \joint detection rate of such events
and to unleash the full potential they have to constrain the \ac{NS}
\ac{EOS}, it will be of paramount importance to have functioning
high-energy gamma-ray observing facilities during the lifespan of
third-generation \ac{GW} detectors.  Finally, other independent
constraints that would reduce our prior on $\rNS$ are expected to
result from ongoing and future missions such as
NICER~\citep{2014SPIE.9144E..20A}, ATHENA~\citep{Motch:2013wfn}, and
eXTP~\citep{Zhang:2016ach}.

\section*{Acknowledgments}
The work presented in this article was supported by Science and
Technology Facilities Council (STFC) grant No.~ST/L000962/1, European
Research Council Consolidator grant No.~647839, and Cardiff University
seedcorn grant AH21101018, as well as the Max Planck Society's Independent 
Research Group programme.
 We acknowledge support from the Amaldi Research Center funded by the MIUR program “Dipartimento di Eccellenza” (CUP:B81I18001170001).
We are grateful for computational resources provided by Cardiff
University, and funded by an STFC grant (ST/I006285/1) supporting UK
Involvement in the Operation of Advanced LIGO.
We thank Stephen Fairhurst and Andrew Williamson for interesting
discussions throughout the genesis of this work.
We also thank Michal Was for his useful comments and input.
N.D.L. acknowledges support from Cardiff University's Leonid Grishchuk
Summer Internship in Gravitational Physics programme.
S.A. acknowledges Stefania Marassi, Silvia Piranomonte, Alessandro
Papitto, Luigi Stella, Enzo Brocato, Viviana Fafone, Valeria Ferrari,
and Cole Miller for useful discussions. S.A. thanks the Cardiff
University School of Physics and Astronomy for the hospitality
received while completing part of this work. S.A. acknowledges
the GRAvitational Wave Inaf TeAm - GRAWITA (P.I. E. Brocato).

%
\bibliography{references}

\begin{thebibliography}{}
\expandafter\ifx\csname natexlab\endcsname\relax\def\natexlab#1{#1}\fi

\bibitem[{Aasi {et~al.}(2015{\natexlab{a}})}]{AdvLIGO}
Aasi, J., {et~al.} 2015{\natexlab{a}}, Class. Quant. Grav., 32, 074001

\bibitem[{Aasi {et~al.}(2015{\natexlab{b}})}]{TheLIGOScientific:2014jea}
---. 2015{\natexlab{b}}, Class. Quantum Grav., 32, 074001

\bibitem[{Abbott {et~al.}(2013)}]{Aasi:2013wya}
Abbott, B.~P., {et~al.} 2013, arXiv:1304.0670, [Living Rev. Rel.19,1(2016)]

\bibitem[{Abbott {et~al.}(2016{\natexlab{a}})}]{GW151226}
---. 2016{\natexlab{a}}, Phys. Rev. Lett., 116, 241103

\bibitem[{Abbott {et~al.}(2016{\natexlab{b}})}]{GW150914}
---. 2016{\natexlab{b}}, Phys. Rev. Lett., 116, 061102

\bibitem[{Abbott {et~al.}(2017{\natexlab{a}})}]{GW170817kn}
---. 2017{\natexlab{a}}, Astrophys. J., 850, L39

\bibitem[{Abbott {et~al.}(2017{\natexlab{b}})}]{GW170817GRB}
---. 2017{\natexlab{b}}, Astrophys. J., 848, L13

\bibitem[{Abbott {et~al.}(2017{\natexlab{c}})}]{GW170104}
---. 2017{\natexlab{c}}, Phys. Rev. Lett., 118, 221101

\bibitem[{Abbott {et~al.}(2017{\natexlab{d}})}]{GW170608}
---. 2017{\natexlab{d}}, Astrophys. J., 851, L35

\bibitem[{Abbott {et~al.}(2017{\natexlab{e}})}]{GW170814}
---. 2017{\natexlab{e}}, Phys. Rev. Lett., 119, 141101

\bibitem[{Abbott {et~al.}(2017{\natexlab{f}})}]{GW170817}
---. 2017{\natexlab{f}}, Phys. Rev. Lett., 119, 161101

\bibitem[{Abbott {et~al.}(2017{\natexlab{g}})}]{GW170817MMA}
---. 2017{\natexlab{g}}, Astrophys. J., 848, L12

\bibitem[{{Abbott} {et~al.}(2018){Abbott}, {Abbott}, {Abbott}, {Acernese},
  {Ackley}, {Adams}, {Adams}, {Addesso}, {Adhikari}, {Adya}, \&
  et~al.}]{GW170817EOS}
{Abbott}, B.~P., {Abbott}, R., {Abbott}, T.~D., {et~al.} 2018, Physical Review
  Letters, 121, 161101

\bibitem[{Abbott {et~al.}(2018)}]{LIGOScientific:2018mvr}
Abbott, B.~P., {et~al.} 2018, arXiv:1811.12907

\bibitem[{{Abbott} {et~al.}(2019){Abbott}, {Abbott}, {Abbott}, {Acernese},
  {Ackley}, {Adams}, {Adams}, {Addesso}, {Adhikari}, {Adya}, \&
  et~al.}]{GW170817PE}
{Abbott}, B.~P., {Abbott}, R., {Abbott}, T.~D., {et~al.} 2019, Physical Review
  X, 9, 011001

\bibitem[{Abernathy {et~al.}(2011)}]{ETdesignstudy}
Abernathy, M., {et~al.} 2011, Einstein gravitational wave Telescope conceptual
  design study. ET-0106C-10, \url{https://tds.ego-gw.it/ql/?c=7954}, ,

\bibitem[{Acernese {et~al.}(2015)}]{AdvVirgo}
Acernese, F., {et~al.} 2015, Class. Quant. Grav., 32, 024001

\bibitem[{{Akmal} {et~al.}(1998){Akmal}, {Pandharipande}, \&
  {Ravenhall}}]{Akmal1998a}
{Akmal}, A., {Pandharipande}, V.~R., \& {Ravenhall}, D.~G. 1998, PhRvC, 58,
  1804

\bibitem[{{Arzoumanian} {et~al.}(2014){Arzoumanian}, {Gendreau}, {Baker},
  {Cazeau}, {Hestnes}, {Kellogg}, {Kenyon}, {Kozon}, {Liu}, {Manthripragada},
  {Markwardt}, {Mitchell}, {Mitchell}, {Monroe}, {Okajima}, {Pollard},
  {Powers}, {Savadkin}, {Winternitz}, {Chen}, {Wright}, {Foster}, {Prigozhin},
  {Remillard}, \& {Doty}}]{2014SPIE.9144E..20A}
{Arzoumanian}, Z., {Gendreau}, K.~C., {Baker}, C.~L., {et~al.} 2014, in
  \procspie, Vol. 9144, Space Telescopes and Instrumentation 2014: Ultraviolet
  to Gamma Ray, 914420

\bibitem[{Aso {et~al.}(2013)Aso, Michimura, Somiya, Ando, Miyakawa, Sekiguchi,
  Tatsumi, \& Yamamoto}]{KAGRA}
Aso, Y., Michimura, Y., Somiya, K., {et~al.} 2013, Phys. Rev., D88, 043007

\bibitem[{Bardeen {et~al.}(1972)Bardeen, Press, \& Teukolsky}]{Bardeen1972}
Bardeen, J.~M., Press, W.~H., \& Teukolsky, S.~A. 1972, Astrophys. J., 178, 347

\bibitem[{{Berger}(2014)}]{Berger2014}
{Berger}, E. 2014, \araa, 52, 43

\bibitem[{{Bildsten} \& {Cutler}(1992)}]{BC92}
{Bildsten}, L., \& {Cutler}, C. 1992, \apj, 400, 175

\bibitem[{{Blandford} \& {Payne}(1982)}]{BP82}
{Blandford}, R.~D., \& {Payne}, D.~G. 1982, \mnras, 199, 883

\bibitem[{{Blandford} \& {Znajek}(1977)}]{BZ77}
{Blandford}, R.~D., \& {Znajek}, R.~L. 1977, \mnras, 179, 433

\bibitem[{{Blinnikov} {et~al.}(1984){Blinnikov}, {Novikov}, {Perevodchikova},
  \& {Polnarev}}]{Blinnikov1984}
{Blinnikov}, S.~I., {Novikov}, I.~D., {Perevodchikova}, T.~V., \& {Polnarev},
  A.~G. 1984, SvAL, 10, 177

\bibitem[{{Bogdanov} {et~al.}(2016){Bogdanov}, {Heinke}, {{\"O}zel}, \&
  {G{\"u}ver}}]{BHOG16}
{Bogdanov}, S., {Heinke}, C.~O., {{\"O}zel}, F., \& {G{\"u}ver}, T. 2016, \apj,
  831, 184

\bibitem[{{Chabanat} {et~al.}(1998){Chabanat}, {Bonche}, {Haensel}, {Meyer}, \&
  {Schaeffer}}]{SLy4}
{Chabanat}, E., {Bonche}, P., {Haensel}, P., {Meyer}, J., \& {Schaeffer}, R.
  1998, Nuclear Physics A, 635, 231

\bibitem[{{Cipolletta} {et~al.}(2015){Cipolletta}, {Cherubini}, {Filippi},
  {Rueda}, \& {Ruffini}}]{Cipolletta2015}
{Cipolletta}, F., {Cherubini}, C., {Filippi}, S., {Rueda}, J.~A., \& {Ruffini},
  R. 2015, \prd, 92, 023007

\bibitem[{Clark {et~al.}(2015)Clark, Evans, Fairhurst, Harry, Macdonald,
  Macleod, Sutton, \& Williamson}]{C15}
Clark, J., Evans, H., Fairhurst, S., {et~al.} 2015, Astrophys. J., 809, 53

\bibitem[{{Demorest} {et~al.}(2010){Demorest}, {Pennucci}, {Ransom}, {Roberts},
  \& {Hessels}}]{Demorest2010}
{Demorest}, P.~B., {Pennucci}, T., {Ransom}, S.~M., {Roberts}, M.~S.~E., \&
  {Hessels}, J.~W.~T. 2010, \nat, 467, 1081

\bibitem[{{Dietrich} {et~al.}(2018){Dietrich}, {Khan}, {Dudi}, {Kapadia},
  {Kumar}, {Nagar}, {Ohme}, {Pannarale}, {Samajdar}, {Bernuzzi}, {Carullo},
  {Del Pozzo}, {Haney}, {Markakis}, {Puerrer}, {Riemenschneider}, {Eka
  Setyawati}, {Tsang}, \& {Van Den Broeck}}]{DK18}
{Dietrich}, T., {Khan}, S., {Dudi}, R., {et~al.} 2018, ArXiv e-prints,
  arXiv:1804.02235

\bibitem[{{Duez} {et~al.}(2010){Duez}, {Foucart}, {Kidder}, {Ott}, \&
  {Teukolsky}}]{DFKOT10}
{Duez}, M.~D., {Foucart}, F., {Kidder}, L.~E., {Ott}, C.~D., \& {Teukolsky},
  S.~A. 2010, Classical and Quantum Gravity, 27, 114106

\bibitem[{{Eichler} {et~al.}(1989){Eichler}, {Livio}, {Piran}, \&
  {Schramm}}]{ELPS89}
{Eichler}, D., {Livio}, M., {Piran}, T., \& {Schramm}, D.~N. 1989, \nat, 340,
  126

\bibitem[{{Engvik} {et~al.}(1996){Engvik}, {Osnes}, {Hjorth-Jensen}, {Bao}, \&
  {Ostgaard}}]{ENG}
{Engvik}, L., {Osnes}, E., {Hjorth-Jensen}, M., {Bao}, G., \& {Ostgaard}, E.
  1996, \apj, 469, 794

\bibitem[{Fairhurst(2018)}]{Fairhurst:2017mvj}
Fairhurst, S. 2018, Class. Quant. Grav., 35, 105002

\bibitem[{Fernández \& Metzger(2016)}]{Fernandez:2015use}
Fernández, R., \& Metzger, B.~D. 2016, Ann. Rev. Nucl. Part. Sci., 66, 23

\bibitem[{{Foucart}(2012)}]{Foucart2012}
{Foucart}, F. 2012, PhRvD, 86, 124007

\bibitem[{{Foucart} {et~al.}(2018){Foucart}, {Hinderer}, \&
  {Nissanke}}]{Foucart2018}
{Foucart}, F., {Hinderer}, T., \& {Nissanke}, S. 2018, ArXiv e-prints,
  arXiv:1807.00011

\bibitem[{Foucart {et~al.}(2013)Foucart, Deaton, Duez, Kidder, MacDonald, Ott,
  Pfeiffer, Scheel, Szilagyi, \& Teukolsky}]{Foucart:2012vn}
Foucart, F., Deaton, M.~B., Duez, M.~D., {et~al.} 2013, Phys. Rev., D87, 084006

\bibitem[{Foucart {et~al.}(2014)Foucart, Deaton, Duez, O'Connor, Ott, Haas,
  Kidder, Pfeiffer, Scheel, \& Szilagyi}]{Foucart:2014nda}
---. 2014, Phys. Rev., D90, 024026

\bibitem[{{Foucart} {et~al.}(2017){Foucart}, {Desai}, {Brege}, {Duez}, {Kasen},
  {Hemberger}, {Kidder}, {Pfeiffer}, \& {Scheel}}]{Foucart2017}
{Foucart}, F., {Desai}, D., {Brege}, W., {et~al.} 2017, Classical and Quantum
  Gravity, 34, 044002

\bibitem[{{Giacomazzo} {et~al.}(2013){Giacomazzo}, {Perna}, {Rezzolla},
  {Troja}, \& {Lazzati}}]{Giacomazzo2013}
{Giacomazzo}, B., {Perna}, R., {Rezzolla}, L., {Troja}, E., \& {Lazzati}, D.
  2013, \apjl, 762, L18

\bibitem[{{Glendenning}(1992)}]{Glendenning1992}
{Glendenning}, N.~K. 1992, \prd, 46, 1274

\bibitem[{{Guillot} {et~al.}(2011){Guillot}, {Rutledge}, \& {Brown}}]{GRB11}
{Guillot}, S., {Rutledge}, R.~E., \& {Brown}, E.~F. 2011, \apj, 732, 88

\bibitem[{{G{\"u}ver} \& {{\"O}zel}(2013)}]{GO13}
{G{\"u}ver}, T., \& {{\"O}zel}, F. 2013, \apjl, 765, L1

\bibitem[{{G{\"u}ver} {et~al.}(2010{\natexlab{a}}){G{\"u}ver}, {{\"O}zel},
  {Cabrera-Lavers}, \& {Wroblewski}}]{GOCW10}
{G{\"u}ver}, T., {{\"O}zel}, F., {Cabrera-Lavers}, A., \& {Wroblewski}, P.
  2010{\natexlab{a}}, \apj, 712, 964

\bibitem[{{G{\"u}ver} {et~al.}(2010{\natexlab{b}}){G{\"u}ver}, {Wroblewski},
  {Camarota}, \& {{\"O}zel}}]{GWCO10}
{G{\"u}ver}, T., {Wroblewski}, P., {Camarota}, L., \& {{\"O}zel}, F.
  2010{\natexlab{b}}, \apj, 719, 1807

\bibitem[{Hannam {et~al.}(2014)Hannam, Schmidt, Boh{\'e}, Haegel, Husa, Ohme,
  Pratten, \& P{\"u}rrer}]{Hannam:2013oca}
Hannam, M., Schmidt, P., Boh{\'e}, A., {et~al.} 2014, Phys. Rev. Lett., 113,
  151101

\bibitem[{{Hawley} \& {Krolik}(2006)}]{HK06}
{Hawley}, J.~F., \& {Krolik}, J.~H. 2006, \apj, 641, 103

\bibitem[{{Heinke} {et~al.}(2006){Heinke}, {Rybicki}, {Narayan}, \&
  {Grindlay}}]{HRNG06}
{Heinke}, C.~O., {Rybicki}, G.~B., {Narayan}, R., \& {Grindlay}, J.~E. 2006,
  \apj, 644, 1090

\bibitem[{{Hinderer} {et~al.}(2016){Hinderer}, {Taracchini}, {Foucart},
  {Buonanno}, {Steinhoff}, {Duez}, {Kidder}, {Pfeiffer}, {Scheel}, {Szilagyi},
  {Hotokezaka}, {Kyutoku}, {Shibata}, \& {Carpenter}}]{HTF16}
{Hinderer}, T., {Taracchini}, A., {Foucart}, F., {et~al.} 2016, Physical Review
  Letters, 116, 181101

\bibitem[{{Hinderer} {et~al.}(2018){Hinderer}, {Nissanke}, {Foucart},
  {Hotokezaka}, {Vincent}, {Kasliwal}, {Schmidt}, {Williamson}, {Nichols},
  {Duez}, {Kidder}, {Pfeiffer}, \& {Scheel}}]{Hinderer2018}
{Hinderer}, T., {Nissanke}, S., {Foucart}, F., {et~al.} 2018, Discerning the
  binary neutron star or neutron star-black hole nature of GW170817 with
  Gravitational Wave and Electromagnetic Measurements, ({\it in preparation}),
  ,

\bibitem[{Husa {et~al.}(2016)Husa, Khan, Hannam, P{\"u}rrer, Ohme, Forteza, \&
  Boh{\'e}}]{Husa:2015iqa}
Husa, S., Khan, S., Hannam, M., {et~al.} 2016, Phys. Rev. D, 93, 044006

\bibitem[{{Iyer} {et~al.}(2011)}]{M1100296}
{Iyer}, B., {et~al.} 2011, {LIGO India}, Tech. Rep. LIGO-M1100296,
  https://dcc.ligo.org/LIGO-M1100296/public

\bibitem[{Kalogera {et~al.}(2019)}]{Kalogera:2019sui}
Kalogera, V., {et~al.} 2019, arXiv:1903.09220

\bibitem[{Kawaguchi {et~al.}(2015)Kawaguchi, Kyutoku, Nakano, Okawa, Shibata,
  \& Taniguchi}]{Kawaguchi:2015bwa}
Kawaguchi, K., Kyutoku, K., Nakano, H., {et~al.} 2015, Phys. Rev., D92, 024014

\bibitem[{{Kawaguchi} {et~al.}(2016){Kawaguchi}, {Kyutoku}, {Shibata}, \&
  {Tanaka}}]{KKST16}
{Kawaguchi}, K., {Kyutoku}, K., {Shibata}, M., \& {Tanaka}, M. 2016, \apj, 825,
  52

\bibitem[{Kawaguchi {et~al.}(2016)Kawaguchi, Kyutoku, Shibata, \&
  Tanaka}]{Kawaguchi:2016ana}
Kawaguchi, K., Kyutoku, K., Shibata, M., \& Tanaka, M. 2016, Astrophys. J.,
  825, 52

\bibitem[{Khan {et~al.}(2016)Khan, Husa, Hannam, Ohme, P{\"u}rrer, Forteza, \&
  Boh{\'e}}]{Khan:2015jqa}
Khan, S., Husa, S., Hannam, M., {et~al.} 2016, Phys. Rev. D, 93, 044007

\bibitem[{{Kokkotas} \& {Schafer}(1995)}]{KS95}
{Kokkotas}, K.~D., \& {Schafer}, G. 1995, \mnras, 275, 301

\bibitem[{Kulkarni(2005)}]{Kulkarni:2005jw}
Kulkarni, S.~R. 2005, arXiv:astro-ph/0510256

\bibitem[{{Kumar} {et~al.}(2017){Kumar}, {P{\"u}rrer}, \& {Pfeiffer}}]{KPP17}
{Kumar}, P., {P{\"u}rrer}, M., \& {Pfeiffer}, H.~P. 2017, \prd, 95, 044039

\bibitem[{{Kyutoku} {et~al.}(2015){Kyutoku}, {Ioka}, {Okawa}, {Shibata}, \&
  {Taniguchi}}]{KI15}
{Kyutoku}, K., {Ioka}, K., {Okawa}, H., {Shibata}, M., \& {Taniguchi}, K. 2015,
  \prd, 92, 044028

\bibitem[{{Kyutoku} {et~al.}(2011){Kyutoku}, {Okawa}, {Shibata}, \&
  {Taniguchi}}]{KOST11}
{Kyutoku}, K., {Okawa}, H., {Shibata}, M., \& {Taniguchi}, K. 2011, \prd, 84,
  064018

\bibitem[{{Kyutoku} {et~al.}(2010){Kyutoku}, {Shibata}, \&
  {Taniguchi}}]{Kyutoku2010}
{Kyutoku}, K., {Shibata}, M., \& {Taniguchi}, K. 2010, PhRvD, 82, 044049

\bibitem[{{Lackey} {et~al.}(2012){Lackey}, {Kyutoku}, {Shibata}, {Brady}, \&
  {Friedman}}]{Lackey2014}
{Lackey}, B.~D., {Kyutoku}, K., {Shibata}, M., {Brady}, P.~R., \& {Friedman},
  J.~L. 2012, PhRvD, 85, 044061

\bibitem[{{Lackey} {et~al.}(2014){Lackey}, {Kyutoku}, {Shibata}, {Brady}, \&
  {Friedman}}]{LK14}
---. 2014, \prd, 89, 043009

\bibitem[{{Lattimer} \& {Prakash}(2016)}]{LP16}
{Lattimer}, J.~M., \& {Prakash}, M. 2016, \physrep, 621, 127

\bibitem[{{Lattimer} {et~al.}(1990){Lattimer}, {Prakash}, {Masak}, \&
  {Yahil}}]{Lattimer1990}
{Lattimer}, J.~M., {Prakash}, M., {Masak}, D., \& {Yahil}, A. 1990, \apj, 355,
  241

\bibitem[{{Leahy} {et~al.}(2008){Leahy}, {Morsink}, \& {Cadeau}}]{LMC08}
{Leahy}, D.~A., {Morsink}, S.~M., \& {Cadeau}, C. 2008, \apj, 672, 1119

\bibitem[{{Leahy} {et~al.}(2011){Leahy}, {Morsink}, \& {Chou}}]{LMC11}
{Leahy}, D.~A., {Morsink}, S.~M., \& {Chou}, Y. 2011, \apj, 742, 17

\bibitem[{{Leahy} {et~al.}(2009){Leahy}, {Morsink}, {Chung}, \&
  {Chou}}]{LMCC09}
{Leahy}, D.~A., {Morsink}, S.~M., {Chung}, Y.-Y., \& {Chou}, Y. 2009, \apj,
  691, 1235

\bibitem[{{Lee} \& {Ramirez-Ruiz}(2007)}]{LRR07}
{Lee}, W.~H., \& {Ramirez-Ruiz}, E. 2007, New Journal of Physics, 9, 17

\bibitem[{{Li} \& {Paczy{\'n}ski}(1998)}]{LP98}
{Li}, L.-X., \& {Paczy{\'n}ski}, B. 1998, \apjl, 507, L59

\bibitem[{\mbox{LIGO Scientific Collaboration, Virgo
  Collaboration}(2017)}]{lalinference_o2}
\mbox{LIGO Scientific Collaboration, Virgo Collaboration}. 2017,
  \mbox{LALSuite},
  \href{https://git.ligo.org/lscsoft/lalsuite/tree/lalinference_o2}{https://git.ligo.org/lscsoft/lalsuite/\\tree/lalinference\_o2},
  GitLab

\bibitem[{{M{\'e}sz{\'a}ros}(2006)}]{Mes06}
{M{\'e}sz{\'a}ros}, P. 2006, Reports on Progress in Physics, 69, 2259

\bibitem[{Meszaros \& Rees(1992)}]{MR92}
Meszaros, P., \& Rees, M.~J. 1992, Astrophys. J., 397, 570

\bibitem[{Metzger(2017)}]{Metzger2017}
Metzger, B.~D. 2017, Living Reviews in Relativity, 20, 3

\bibitem[{{Metzger} \& {Berger}(2012)}]{MB12}
{Metzger}, B.~D., \& {Berger}, E. 2012, \apj, 746, 48

\bibitem[{Metzger {et~al.}(2010)Metzger, Mart\'inez-Pinedo, Darbha, Quataert,
  Arcones, Kasen, Thomas, Nugent, Panov, \& Zinner}]{Metzger10}
Metzger, B.~D., Mart\'inez-Pinedo, G., Darbha, S., {et~al.} 2010, Monthly
  Notices of the Royal Astronomical Society, 406, 2650

\bibitem[{{Morsink} \& {Leahy}(2011)}]{ML11}
{Morsink}, S.~M., \& {Leahy}, D.~A. 2011, \apj, 726, 56

\bibitem[{Motch {et~al.}(2013)}]{Motch:2013wfn}
Motch, C., {et~al.} 2013, arXiv:1306.2334

\bibitem[{{M{\"u}ller} \& {Serot}(1996)}]{MS}
{M{\"u}ller}, H., \& {Serot}, B.~D. 1996, Nuclear Physics A, 606, 508

\bibitem[{{M{\"u}ther} {et~al.}(1987){M{\"u}ther}, {Prakash}, \&
  {Ainsworth}}]{MPA}
{M{\"u}ther}, H., {Prakash}, M., \& {Ainsworth}, T.~L. 1987, Physics Letters B,
  199, 469

\bibitem[{Nakar(2007)}]{Nakar:2007yr}
Nakar, E. 2007, PhR, 442, 166

\bibitem[{{Narayan} {et~al.}(1992){Narayan}, {Paczynski}, \& {Piran}}]{NPP92}
{Narayan}, R., {Paczynski}, B., \& {Piran}, T. 1992, \apjl, 395, L83

\bibitem[{Nissanke {et~al.}(2013)Nissanke, Kasliwal, \&
  Georgieva}]{Nissanke:2012dj}
Nissanke, S., Kasliwal, M., \& Georgieva, A. 2013, Astrophys. J., 767, 124

\bibitem[{{Oppenheimer} \& {Volkoff}(1939)}]{OV39}
{Oppenheimer}, J.~R., \& {Volkoff}, G.~M. 1939, Physical Review, 55, 374

\bibitem[{{{\"O}zel} \& {Freire}(2016)}]{OF16}
{{\"O}zel}, F., \& {Freire}, P. 2016, \araa, 54, 401

\bibitem[{{{\"O}zel} {et~al.}(2012){{\"O}zel}, {Gould}, \& {G{\"u}ver}}]{OGG12}
{{\"O}zel}, F., {Gould}, A., \& {G{\"u}ver}, T. 2012, \apj, 748, 5

\bibitem[{{{\"O}zel} {et~al.}(2009){{\"O}zel}, {G{\"u}ver}, \&
  {Psaltis}}]{OGP09}
{{\"O}zel}, F., {G{\"u}ver}, T., \& {Psaltis}, D. 2009, \apj, 693, 1775

\bibitem[{{Paczynski}(1986)}]{Paczynski1986}
{Paczynski}, B. 1986, \apjl, 308, L43

\bibitem[{{Paczynski}(1991)}]{P91}
---. 1991, \actaa, 41, 257

\bibitem[{Pannarale {et~al.}(2015{\natexlab{a}})Pannarale, Berti, Kyutoku,
  Lackey, \& Shibata}]{Pannarale2015b}
Pannarale, F., Berti, E., Kyutoku, K., Lackey, B.~D., \& Shibata, M.
  2015{\natexlab{a}}, Phys. Rev., D92, 084050

\bibitem[{Pannarale {et~al.}(2015{\natexlab{b}})Pannarale, Berti, Kyutoku,
  Lackey, \& Shibata}]{Pannarale2015a}
---. 2015{\natexlab{b}}, Phys. Rev., D92, 081504

\bibitem[{Pannarale {et~al.}(2013)Pannarale, Berti, Kyutoku, \&
  Shibata}]{Pannarale2013a}
Pannarale, F., Berti, E., Kyutoku, K., \& Shibata, M. 2013, Phys. Rev., D88,
  084011

\bibitem[{Pannarale \& Ohme(2014)}]{Pannarale:2014rea}
Pannarale, F., \& Ohme, F. 2014, Astrophys. J., 791, L7

\bibitem[{{Pannarale} {et~al.}(2011){Pannarale}, {Rezzolla}, {Ohme}, \&
  {Read}}]{PannaraleRezzolla2011}
{Pannarale}, F., {Rezzolla}, L., {Ohme}, F., \& {Read}, J.~S. 2011, \prd, 84,
  104017

\bibitem[{Pannarale {et~al.}(2011)Pannarale, Tonita, \&
  Rezzolla}]{Pannarale2010}
Pannarale, F., Tonita, A., \& Rezzolla, L. 2011, ApJ, 727, 95

\bibitem[{{Parfrey} {et~al.}(2015){Parfrey}, {Giannios}, \&
  {Beloborodov}}]{PGB15}
{Parfrey}, K., {Giannios}, D., \& {Beloborodov}, A.~M. 2015, \mnras, 446, L61

\bibitem[{{Poutanen} \& {Gierli{\'n}ski}(2003)}]{PG03}
{Poutanen}, J., \& {Gierli{\'n}ski}, M. 2003, \mnras, 343, 1301

\bibitem[{Punturo {et~al.}(2010)}]{Punturo:2010zz}
Punturo, M., {et~al.} 2010, Class. Quant. Grav., 27, 194002

\bibitem[{Regimbau {et~al.}(2015)Regimbau, Siellez, Meacher, Gendre, \&
  Boër}]{Regimbau:2014nxa}
Regimbau, T., Siellez, K., Meacher, D., Gendre, B., \& Boër, M. 2015,
  Astrophys. J., 799, 69

\bibitem[{{Sari} {et~al.}(1999){Sari}, {Piran}, \& {Halpern}}]{SPH99}
{Sari}, R., {Piran}, T., \& {Halpern}, J.~P. 1999, \apjl, 519, L17

\bibitem[{Sathyaprakash {et~al.}(2019)}]{Sathyaprakash:2019rom}
Sathyaprakash, B.~S., {et~al.} 2019, arXiv:1903.09277

\bibitem[{{Setiawan} {et~al.}(2004){Setiawan}, {Ruffert}, \& {Janka}}]{SRJ04}
{Setiawan}, S., {Ruffert}, M., \& {Janka}, H.-T. 2004, \mnras, 352, 753

\bibitem[{Shibata {et~al.}(2009)Shibata, Kyutoku, Yamamoto, \&
  Taniguchi}]{Shibata:2009cn}
Shibata, M., Kyutoku, K., Yamamoto, T., \& Taniguchi, K. 2009, Phys. Rev., D79,
  044030, [Erratum: Phys. Rev.D85,127502(2012)]

\bibitem[{Smith {et~al.}(2016)Smith, Field, Blackburn, Haster, P{\"u}rrer,
  Raymond, \& Schmidt}]{Smith:2016qas}
Smith, R., Field, S.~E., Blackburn, K., {et~al.} 2016, Phys. Rev. D, 94, 044031

\bibitem[{{Tolman}(1939)}]{T39}
{Tolman}, R.~C. 1939, Physical Review, 55, 364

\bibitem[{Unnikrishnan(2013)}]{LigoIndia}
Unnikrishnan, C.~S. 2013, Int. J. Mod. Phys., D22, 1341010

\bibitem[{{Vallisneri}(2000)}]{V00}
{Vallisneri}, M. 2000, Physical Review Letters, 84, 3519

\bibitem[{{van Paradijs}(1979)}]{V79}
{van Paradijs}, J. 1979, \apj, 234, 609

\bibitem[{{Veitch} {et~al.}(2015){Veitch}, {Raymond}, {Farr}, {Farr}, {Graff},
  {Vitale}, {Aylott}, {Blackburn}, {Christensen}, {Coughlin}, {Del Pozzo},
  {Feroz}, {Gair}, {Haster}, {Kalogera}, {Littenberg}, {Mandel},
  {O'Shaughnessy}, {Pitkin}, {Rodriguez}, {R{\"o}ver}, {Sidery}, {Smith}, {Van
  Der Sluys}, {Vecchio}, {Vousden}, \& {Wade}}]{Veitch2015}
{Veitch}, J., {Raymond}, V., {Farr}, B., {et~al.} 2015, \prd, 91, 042003

\bibitem[{Wanderman \& Piran(2015)}]{Wanderman:2014eza}
Wanderman, D., \& Piran, T. 2015, Mon. Not. Roy. Astron. Soc., 448, 3026

\bibitem[{{Webb} \& {Barret}(2007)}]{WB07}
{Webb}, N.~A., \& {Barret}, D. 2007, \apj, 671, 727

\bibitem[{{Wiringa} {et~al.}(1988){Wiringa}, {Fiks}, \& {Fabrocini}}]{WFF1}
{Wiringa}, R.~B., {Fiks}, V., \& {Fabrocini}, A. 1988, \prc, 38, 1010

\bibitem[{{Yagi} \& {Yunes}(2017)}]{YY17}
{Yagi}, K., \& {Yunes}, N. 2017, \physrep, 681, 1

\bibitem[{{Zalamea} \& {Beloborodov}(2011)}]{ZB11}
{Zalamea}, I., \& {Beloborodov}, A.~M. 2011, \mnras, 410, 2302

\bibitem[{{Zhang} {et~al.}(2007){Zhang}, {Liang}, {Page}, {Grupe}, {Zhang},
  {Barthelmy}, {Burrows}, {Campana}, {Chincarini}, {Gehrels}, {Kobayashi},
  {M{\'e}sz{\'a}ros}, {Moretti}, {Nousek}, {O'Brien}, {Osborne}, {Roming},
  {Sakamoto}, {Schady}, \& {Willingale}}]{Zhang2007}
{Zhang}, B., {Liang}, E., {Page}, K.~L., {et~al.} 2007, \apj, 655, 989

\bibitem[{Zhang {et~al.}(2016)}]{Zhang:2016ach}
Zhang, S.~N., {et~al.} 2016, Proc. SPIE Int. Soc. Opt. Eng., 9905, 99051Q

\end{thebibliography}

\appendix
\renewcommand*\thetable{\Alph{section}A.\arabic{table}}


\begin{table*}[ht]
\centering
\setlength{\tabcolsep}{10pt}
	\begin{tabular}{cccccc}
		\toprule
    	\multirow{2}{*}{Label} & \multirow{2}{*}{\ac{SNR}}&$\mNS$ & $\mBH$ & \multirow{2}{*}{$\spin$} & \multirow{2}{*}{$\rm q$}\\
         & & $[M_\odot]$ & $[M_\odot]$\\
		\midrule
		\multirow{2}{*}{\texttt{m484chi048(H/L)}} & 10 & 1.25 - 2.46 & 2.57 - 5.47& 0.14 - 0.81 & 1.05 - 4.31 \\
		& 30 & 1.28 - 1.94 & 3.19 - 5.26 & 0.23 - 0.68 & 1.64 - 4.11\\
        
        \texttt{m484chi080(H/L)} & 30 & 1.19 - 2.29 & 2.66 - 5.72 & 0.65 - 0.88 & 1.16 - 4.79  \\
        \texttt{m100chi070(H/L)} & 30 & 1.32 - 1.56 & 8.30 - 10.44 & 0.67 - 0.76 & 5.32 - 7.89 \\
		\bottomrule
	\end{tabular}
	\caption{90\% credible intervals of binary system properties obtained by the parameter estimation of the \ac{GW} signal injections used in our study.}
	\label{tab:fullposterior}
\end{table*}

\begin{table*}[ht]
\centering
\setlength{\tabcolsep}{10pt}
	\begin{tabular}{cccccc}
		\toprule
    	\multirow{2}{*}{Label} & \multirow{2}{*}{\ac{SNR}}&$\mNS$ & $\mBH$ & \multirow{2}{*}{$\spin$} & \multirow{2}{*}{$\rm q$}\\
         & & $[M_\odot]$ & $[M_\odot]$\\
		\midrule
		\multirow{2}{*}{\texttt{m484chi048(H/L)}} & 10 & 1.20 - 2.12 & 2.57 - 5.47& 0.14 - 0.81 & 1.45 - 4.83 \\
		& 30 & 1.29 - 1.83 & 3.05 - 5.84 & 0.38 - 0.68 & 1.86 - 4.07\\
        
        \texttt{m484chi080(H/L)} & 30 & 1.13 - 1.89 & 3.30 - 6.17 & 0.76 - 0.88 & 1.74 - 5.46  \\
        \texttt{m100chi070(H/L)} & 30 & 1.32 - 1.53 & 8.52 - 10.46 & 0.68 - 0.72 & 5.56 - 7.90 \\
		\bottomrule
	\end{tabular}
	\caption{Same as Table \ref{tab:fullposterior}, but obtained from the posterior probability distribution after it is pruned in order to select only parameters compatible with \nsbh systems.}
	\label{tab:nsbh_posterior}
\end{table*}

\end{document}